\documentclass[vecphys]{svmult}
\usepackage{graphicx}       
\usepackage{multicol}       
\usepackage[bottom]{footmisc}
\makeindex             
\newcommand{\newc}{\newcommand}
\newc{\ra}{\rightarrow}
\newc{\lra}{\leftrightarrow}
\newc{\beq}{\begin{equation}}
\newc{\eeq}{\end{equation}}
\newc{\barr}{\begin{eqnarray}}
\newc{\earr}{\end{eqnarray}}

\begin{document}

\title*{On The Direct Detection of Dark  Matter-\\ Exploring all the signatures of the neutralino-nucleus interaction}
\titlerunning{On The Direct Detection of Dark  Matter}
\author{J.D. Vergados}
\authorrunning{J.D. Vergados} 
\institute{Physics Department, University of Ioannina, Gr 451 10, Ioannina, Greece
\texttt{vergados@cc.uoi.gr}}
%
%
\maketitle
\begin{abstract}
 Various issues related to the direct detection of supersymmetric dark matter are reviewed. Such are: 1) Construction of  supersymmetric models with a number of parameters, which are constrained from the data at low energies as well as cosmological observations. 2) A model for the nucleon, in particular the dependence on the nucleon cross section on quarks other than u and d. 3) A nuclear model, i.e.  the nuclear form factor for the scalar interaction and the spin response function for the axial current. 4) Information about the density and the velocity distribution of the neutralino (halo model). Using the present experimental limits on the rates and proper inputs in 3)-4) we derive constraints in the nucleon cross section, which involves 1)-2). Since the expected event rates are extremely low we consider some additional signatures of the neutralino nucleus interaction, such as the periodic behavior of the rates due to the motion of Earth (modulation effect), which, unfortunately, is characterized by a small amplitude. This leads us to examine the possibility of suggesting directional experiments, which  measure not only the energy of the recoiling nuclei but their direction as well. In these, albeit hard, experiments one can exploit two very characteristic signatures: a)large asymmetries and b) interesting modulation patterns. Furthermore we  extended our study to include evaluation of the rates for other than recoil searches such as: i) Transitions to excited states, ii) Detection of recoiling electrons produced during the neutralino-nucleus interaction and iii) Observation of hard X-rays following the de-excitation of the ionized atom. 
\end{abstract}
\section{Introduction}
\label{sec:1}
The combined MAXIMA-1 \cite{MAXIMA-1}, BOOMERANG \cite{BOOMERANG},
DASI \cite{DASI}, COBE/DMR Cosmic Microwave Background (CMB)
observations \cite{COBE}, the recent WMAP data \cite{SPERGEL} and
SDSS
 \cite{SDSS} imply that the
Universe is flat \cite{flat01} and
 and that most of the matter in
the Universe is dark, i.e. exotic.
  $$ \Omega_b=0.044\pm 0.04,
\Omega_m=0.27\pm 0.04,  \Omega_{\Lambda}=0.69\pm0.08$$
 for baryonic matter , cold dark matter and dark energy
respectively. An analysis of a combination of SDSS and WMAP data
yields \cite{SDSS} $\Omega_m\approx0.30\pm0.04(1\sigma)$. Crudely
speaking and easy to remember
$$\Omega_b\approx 0.05, \Omega _{CDM}\approx 0.30, \Omega_{\Lambda}\approx 0.65$$
Additional indirect information
comes from  the rotational curves \cite{Jung}. The rotational velocity of an object increases so long
is surrounded but matter. Once outside matter the velocity of rotation drops as the square root  of the 
distance. Such observations are not possible in our own galaxy. The observations of other galaxies, 
similar to our own, indicate that the rotational velocities of objects outside the luminous matter
do not drop. So there must be a halo of dark matter out there.
Since the non exotic component cannot exceed $40\%$ of the CDM
~\cite {Benne}, there is room for exotic WIMP's (Weakly
Interacting Massive Particles).\\
  In fact the DAMA experiment ~\cite {BERNA2} has claimed the observation of one signal in direct
detection of a WIMP, which with better statistics has subsequently
been interpreted as a modulation signal \cite{BERNA1}.These  data,
however, if they are due to the coherent process, are not
consistent with other recent experiments, see e.g. EDELWEISS and
CDMS \cite{EDELWEISS}. It could still be interpreted as due to the
spin cross section, but with a new interpretation of the extracted
nucleon cross section.
The above developments are in line with particle physics considerations. Thus,
in the currently favored supersymmetric (SUSY)
extensions of the standard model, the most natural WIMP candidate is the LSP,
i.e. the lightest supersymmetric particle. In the most favored scenarios the
LSP can be simply described as a Majorana fermion, a linear 
combination of the neutral components of the gauginos and Higgsinos
\cite{Jung}-\cite{Hab-Ka}. 

 Since this particle is expected to be very massive, $m_{\chi} \geq 30 GeV$, and
extremely non relativistic with average kinetic energy $T \leq 100 KeV$,
it can be directly detected mainly via the recoiling
of a nucleus (A,Z) in the elastic scattering process:
\begin{equation}
\chi \, +\, (A,Z) \, \to \, \chi \,  + \, (A,Z)^* 
\end{equation}
($\chi$ denotes the LSP). In order to compute the event rate one needs
the following ingredients:
\begin{enumerate}
\item An effective Lagrangian at the elementary particle 
(quark) level obtained in the framework of supersymmetry 
~\cite{Jung}, \cite{ref2} and \cite{Hab-Ka}.
 One starts with   
representative input in the restricted SUSY parameter space as described in
the literature, e.g. Ellis {\it et al} \cite{EOSS04}, Bottino {\it et al}, 
Kane {\it et al.}, Castano {\it et al.} and Arnowitt {\it et al} \cite{ref2} as well as elsewhere
\cite{GOODWIT}-\cite{UK01}.
We will not, however, elaborate on how one gets the needed parameters from
supersymmetry, since this topic will be covered by another lecture in this school by professor 
Lahanas. 
For the reader's convenience, however, we will give a description  in sec. \ref{sec:diagrams} of the basic 
SUSY ingredients needed to calculate LSP-nucleus scattering
 cross section. 
 Our own SUSY input parameters can be found elsewhere \cite{Gomez}.
\item A procedure in going from the quark to the nucleon level, i.e. a quark 
model for the nucleon. The results depend crucially on the content of the
nucleon in quarks other than u and d. This is particularly true for the scalar
couplings as well as the isoscalar axial coupling ~\cite{Dree}-\cite{Chen}. Such topics will be discussed in sec.
\ref{sec:nuc}.
\item computation of  the relevant nuclear matrix elements~\cite{Ress}-\cite{SUHONEN03}
using as reliable as possible many body nuclear wave functions.
By putting as accurate nuclear physics input as possible, 
one will be able to constrain the SUSY parameters as much as possible.
The situation is a bit simpler in the case of the scalar coupling, in which
case one only needs the nuclear form factor.
\item  Convolution with the LSP velocity Distribution.
To this end we will consider here  Maxwell-Boltzmann \cite {Jung} (MB) velocity distributions, with an upper velocity cut off put in by hand. Other distributions are possible, such as
  non symmetric ones,  like those 
of Drukier \cite {Druk} and Green \cite{GREEN02}, or non isothermal ones, e.g. those arising from late in-fall of   
dark matter into our galaxy, like  Sikivie's caustic rings \cite {SIKIVIE}. In any event in a proper treatment the velocity distribution ought to be consistent with the dark matter density in the context of the Eddington theory \cite{OWVER}.
\end{enumerate}

After this we will specialize our study in the case of the nucleus $^{127}I$, 
which is one of the most popular targets
\cite{BERNA2}, \cite {KVdubna}.

 Since the expected rates are extremely low or even undetectable
with present techniques, one would like to exploit the characteristic
signatures provided by the reaction. Such are: 
\begin{enumerate}
\item The modulation
 effect, i.e the dependence of the event rate on the velocity of
the Earth 
\item The directional event rate, which depends on the
 velocity of the sun around the galaxy as well as the the velocity
of the Earth. 
has recently begun to appear feasible by the planned experiments
\cite {UKDMC},\cite{DRIFT}.
\item Detection of other than nuclear recoils, such as
\begin{itemize}
\item Detection of $\gamma$ rays following nuclear de-excitation, whenever possible \cite{eji93},\cite{VQS04}.
\item Detection of ionization electrons produced directly in the LSP-nucleus collisions \cite{VE05},\cite{MVE05}. 
\item Observations of hard X-rays produced\cite{EMV05}, when the inner shell electron holes produced as above are filled.
\end{itemize}
\end{enumerate}

 In all calculations we will, of course, include an appropriate 
nuclear form factor and take into account
the influence on the rates of the detector energy cut off.
We will present our results a function of the
LSP mass, $m_{\chi}$, in a way
which can be easily understood by the experimentalists.


\section{The Nature of the LSP}
\label{sec:LSP}

Before proceeding with the
construction of the effective Lagrangian we will briefly discuss 
the nature of the
lightest supersymmetric particle (LSP) focusing on those ingredients
which are of interest to dark matter.

In currently favorable supergravity models the LSP is a linear
combination~\cite{Jung} of the neutral four fermions 
${\tilde B}, {\tilde W}_3, {\tilde H}_1$ and ${\tilde H}_2$ 
which are the supersymmetric partners of the gauge bosons $B_\mu$ and
$W^3_\mu$ and the Higgs scalars
$H_1$ and $H_2$. Admixtures of s-neutrinos are expected to be negligible.

In the above basis the mass-matrix takes the form~\cite{Jung,Hab-Ka} 
\beq
  \left(\begin{array}{cccc}M_1 & 0 & -m_z c_\beta s_w & m_z s_\beta s_w \\
 0 & M_2 & m_z c_\beta c_w & -m_z s_\beta c_w \\
-m_z c_\beta s_w & m_z c_\beta c_w & 0 & -\mu \\
m_z s_\beta s_w & -m_z c_\beta c_w & -\mu & 0 \end{array}\right)
\label{eq:eg 4}
\eeq

In the above expressions $c_W = cos \theta_W$,  
$s_W = sin\theta_W$, $c_\beta = cos\beta$, $s_\beta = sin\beta$,
where $tan\beta =\langle \upsilon_2\rangle/\langle\upsilon_1\rangle$ is the
ratio of the vacuum expectation values of the Higgs scalars $H_2$ and
$H_1$. $\mu$ is a dimensionful coupling constant which is not specified by the
theory (not even its sign). 

By diagonalizing the above matrix we obtain a set of eigenvalues $m_j$ and
the diagonalizing matrix $C_{ij}$ as follows
\beq
\left(\begin{array}{c}{\tilde B}_R\\
{\tilde W}_{3R}\\
{\tilde H}_{1R} \\
 {\tilde H}_{2R}\end{array}\right) = (C^R_{ij})
\left(\begin{array}{c} \chi_{1R} \\
\chi_{2R} \\   \chi_{3R}\\
\chi_{4R}\end{array}\right)  \qquad
\left(\begin{array}{c} {\tilde B}_L \\
{\tilde W}_{2L} \\   {\tilde H}_{1L} \\
{\tilde H}_{2L} \end{array}\right) =  \left(C_{ij} \right)
\left(\begin{array}{c} \chi_{1L} \\
\chi_{2L} \\   \chi_{3L} \\
\chi_{4L} \end{array}\right) 
\label{eq:eg.8}
\eeq

\noindent
with $C^R_{ij}=C^*_{ij}e^{i \lambda_j}$
The phases are $\lambda_i=0, \pi$ depending on the sign of the eigenmass.

Another possibility to express the above results in photino-zino 
basis ${\tilde \gamma}, {\tilde Z}$ via
\beq
{\tilde W}_3 =sin \theta_W {\tilde \gamma}
 -cos \theta_W {\tilde Z}~~,~~
{\tilde B}_0 = cos \theta_W {\tilde \gamma}
 +sin \theta_W {\tilde Z} \label{eq:eg9}
\eeq
In the absence of supersymmetry breaking $(M_1=M_2=M$ and $\mu=0)$ the
photino is one of the eigenstates with mass $M$. One of the remaining
eigenstates has a zero eigenvalue and is a linear combination of ${\tilde
H}_1$ and ${\tilde H}_2$ with mixing  $sin \beta$. In the presence
of SUSY breaking terms the ${\tilde B}, {\tilde W}_3$ basis is superior
since the lowest eigenstate $\chi_1$ or LSP is primarily ${\tilde B}$. From
our point of view the most important parameters are the mass $m_x$ of LSP
and the mixing $C_{j1}, j=1,2,3,4$ which yield the $\chi_1$ content of the
initial basis states.\\
We are now in a position to find the interaction of $\chi_1$ with matter. 
We distinguish three possibilities involving Z-exchange, s-quark exchange and
Higgs exchange.

\section{The Feynman Diagrams Entering the Direct Detection of LSP.} 
\label{sec:diagrams}
 The diagrams involve Z-exchange, s-quark exchange and Higgs exchange.
\subsection{The Z-exchange contribution.}
\label{sec:Z-exc}
\begin{figure}
\begin{center}
\includegraphics[height=4cm]{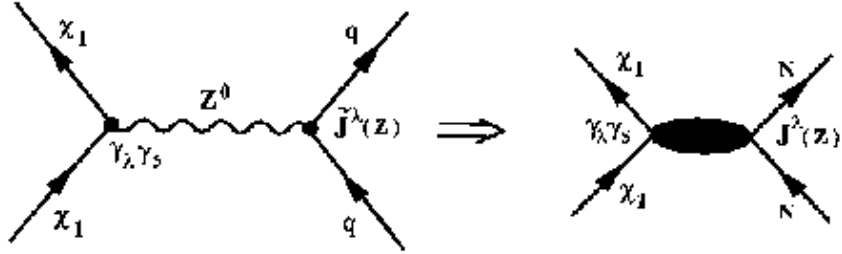}
\caption{ The LSP-quark interaction mediated by Z-exchange.
 \label{LSPZ} }
\end{center}
\end{figure}

This can arise from the interaction of Higgsinos with $Z$ (see Fig. \ref{LSPZ}) which can be read
from Eq. C86 of Ref.~\cite{Hab-Ka} 
\beq
{\it L} = \frac{g}{cos \theta_W} \frac{1}{4} [{\tilde H}_{1R}
\gamma_{\mu}{\tilde H}_{1R} -{\tilde H}_{1L}\gamma_{\mu} {\tilde H}_{1L} -
({\tilde H}_{2R}\gamma_{\mu}{\tilde H}_{2R}  -{\tilde H}_{2L}\gamma_{\mu}{\tilde
H}_{2L})]Z^{\mu}
 \label{eq:eg 10}
\eeq
Using Eq. (\ref{eq:eg.8}) and the fact that for Majorana particles 
${\bar \chi} \gamma_{\mu} \chi = 0$, we obtain 
\beq
{\it L} = \frac {g}{cos \theta_W} \frac {1}{4} (|C_{31}|^2
-|C_{41}|^2) {\bar \chi}_1\gamma_{\mu} \gamma_5 \chi_1 Z^{\mu}
 \label{eq:eg 11}
\eeq
which leads to the effective 4-fermion interaction 
\beq
{\it L}_{eff} = \frac {g}{cos \theta_W} \frac {1}{4} 2(|C_{31}|^2
-|C_{41}|^2)  (- \frac {g}{2cos \theta_W} \frac {1}{q^2 -m^2_Z}
 {\bar \chi}_1\gamma^{\mu} \gamma_5 \chi_1)J^Z_\mu
 \label{eq:eg 12}
\eeq
where the extra factor of 2 comes from the Majorana nature of
$\chi_1$. The neutral hadronic current $J^Z_\lambda$ is given by
\beq
J^Z_{\lambda} = - {\bar q} \gamma_{\lambda} \{ \frac {1}{3} sin^2 \theta_W -
\Big[ \,\frac {1}{2} (1-\gamma_5) - sin^2 \theta_W \Big]\tau_3 \} q  
  \label{eq:eg 13}
\eeq
at the nucleon level it can be written as
\beq
{\tilde J}_{\lambda}^Z = -{\bar N} \gamma_{\lambda} \{ \, sin^2 \theta_W 
-g_V (\frac{1}{2} - sin^2\theta_W)\tau_3 + \frac{1}{2} g_A \gamma_5 \tau_3
\} N
  \label{eq:eg 14}
\eeq
Thus we can write
\beq
{\it L}_{eff} = - \frac {G_F}{\sqrt 2} ({\bar \chi}_1 \gamma^{\lambda}
\gamma^5 \chi_1) J_{\lambda}(Z)
 \label{eq:eg 15}
\eeq 
where
\beq
J_{\lambda}(Z) = {\bar N} \gamma_{\lambda} [f^0_V(Z) + f^1_V(Z) \tau_3
+  f^0_A(Z) \gamma_5 + f^1_A(Z) \gamma_5  \tau_3] N
\label{eq:eg 16}
\eeq
and
\barr
f^0_V(Z) &=& 2(|C_{31}|^2 -|C_{41}|^2) \frac {m^2_Z}{m^2_Z - q^2} sin^2
\theta_W \\ 
f^1_V(Z) &=& - 2(|C_{31}|^2 -|C_{41}|^2) \frac
{m^2_Z}{m^2_Z - q^2}g_V (\frac {1}{2} - sin^2 \theta_W) \\
f^0_A (Z)&=&0 ~~,~~ 
f^1_A (Z) = 2(|C_{31}|^2 -|C_{41}|^2) \frac {m^2_Z}{m^2_Z - q^2} \,\, \frac
{1}{2} g_A \label{eq:eg 13d}
\earr
with $g_V=1.0$ and $g_A = 1.24$. We can easily see that
\beq
f^1_{V}(Z)/ f^0_{V}(Z) = -g_V ( \frac {1}{2sin^2 \theta_W} - 1 ) \simeq
- 1.15  \nonumber
\eeq
Note that the suppression of this Z-exchange interaction compared to 
the ordinary
neutral current interactions arises from the smallness of the mixing
$C_{31}$ and $C_{41}$, a consequence of the fact that the Higgsinos are
normally quite a bit heavier than the gauginos.  Furthermore, the two
Higgsinos tend to cancel each other.

We should also mention that the vector contribution, the time component of which can lead to coherence,
contributes only to order $\upsilon /c\approx 10^{-3}$ due to the Majorana nature of the LSP. Thus to leading order
only the axial current can contribute to the direct detection of the neutralino.

\subsection{The $s$-quark Mediated Interaction }
\label{sec:sq-exc}
\begin{figure}
\begin{center}
\includegraphics[height=8cm]{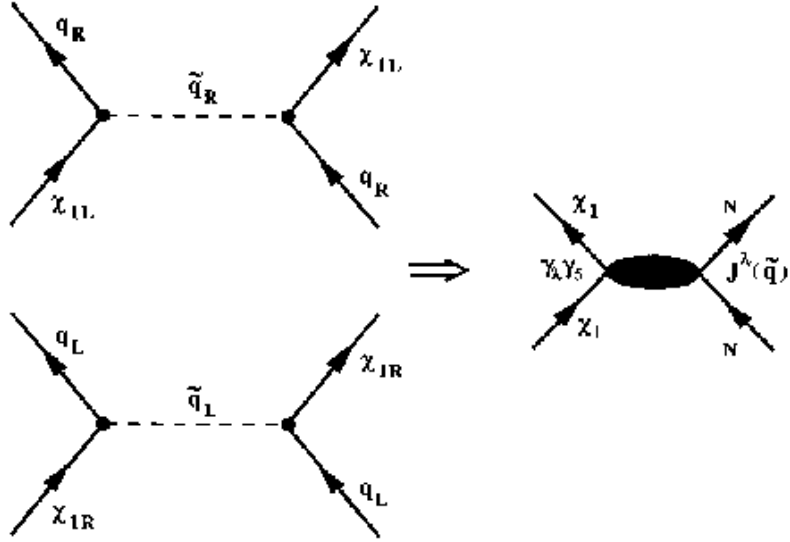}
\hspace{1.0cm}
\caption{ The LSP-quark interaction mediated by s-quark exchange.
 \label{LSPSQ} }
\end{center}
\end{figure}

The other interesting possibility arises from the other two components of
$\chi_1$, namely ${\tilde B}$ and ${\tilde W}_3$. Their corresponding
couplings to $s$-quarks (see Fig. \ref{LSPSQ}) can be read from the appendix C4 of Ref.~\cite{Hab-Ka}
They are
\barr
{\it L}_{eff} &=& -g \sqrt {2} \{{\bar q}_L [T_3 {\tilde W}_{3R} 
- tan \theta_W (T_3 -Q) {\tilde B}_R ] {\tilde q}_L \nonumber \\
&-& tan  \theta_W {\bar q}_R Q {\tilde B}_L {\tilde q}_R\} + HC
 \label{eq:eg 17}
\earr
where ${\tilde q}$ are the scalar quarks (SUSY partners of quarks). A
summation over all quark flavors is understood. Using Eq. (\ref{eq:eg.8}) we
can write the above equation in the $\chi_i$ basis. Of interest to us here
is the part
\barr 
{\it L}_{eff} &=& g \sqrt {2} \{(tan \theta_W (T_3 -Q) C^R_{11} -
T_3 C^R_{21}) {\tilde q}_L \chi_{1R} {\tilde q}_L \nonumber \\
&+& tan  \theta_W C_{11} Q {\bar q}_R \chi_{1L} {\tilde q}_R\} 
\label{eq:eg.18}
\earr
The above interaction is almost diagonal in the quark flavor. There exists,
however, mixing between the s-quarks ${\tilde q}_L$ and ${\tilde q}_R$
(of the same flavor) i.e.
\begin{equation}
{\tilde q}_L = cos \theta_{{\tilde q}}{\tilde q}_1 
+ sin \theta_{{\tilde q}}{\tilde q}_2 ~,~  
{\tilde q}_R = -sin \theta_{{\tilde q}}{\tilde q}_1 
+ cos \theta_{{\tilde q}}{\tilde q}_2   
\end{equation}
with
\begin{equation}
tan 2\theta_{{\tilde u}} =  \frac {m_u(A+\mu cot \beta)} 
{m^2_{u_L} -m^2_{{\tilde u}_R} + m^2_z cos2 \beta/2}~,~
tan 2\theta_{{\tilde d}} =  \frac {m_d(A+\mu tan \beta)} 
{m^2_{d_R} -m^2_{{\tilde d}_R} + m^2_Z cos2 \beta/2}
\end{equation}
Thus Eq. (\ref{eq:eg.18}) becomes
\barr
{\it L}_{eff} &=& g \sqrt{2} \left\{ [ B_L cos\theta_{{\tilde q}} \right.
{\bar q}_L \chi_{1R} -B_R sin\theta_{{\tilde q}}{\bar q}_R \chi_{1L}]
{\tilde q}_1 \nonumber \\
&+& [ B_L sin \theta_{{\tilde q}} {\bar q}_L \chi_{1R} + 
B_R cos\theta_{{\tilde q}} {\bar q}_R \chi_{1L}] \left.
{\tilde q}_2 \right \} \nonumber   
\earr
with
\begin{center}
$B_L(q) = -\frac{1}{6} C^R_{11}tan\theta_{\omega}-\frac{1}{2} C^R_{21},
\,\,\,  q=u \ \ (charge \ \ 2/3)$   
\end{center}
\begin{center}
$B_L(q) = -\frac{1}{6} C^R_{11} tan\theta_{\omega} + \frac{1}{2} C^R_{21},
\ \ \ q=d \ \ (charge \ \ -1/3)$   
\end{center}
\begin{center}
$B_R(q) = \frac {2}{3} tan \theta_{\omega}  C_{11}, \ \ \  q=u \ \ (charge \ \ 2/3)$   
\end{center}
\begin{center}
$B_R(q) = - \frac {1}{3} tan \theta_{\omega}  C_{11}, \ \ \  q=d\ \ (charge \ \ -1/3)$   
\end{center}
The effective four fermion interaction takes the form
\barr
\lefteqn{{\it L}_{eff} = (g \sqrt{2})^2  \{(B_L cos \theta_{\tilde q} {\bar q}_L
\chi_{1R} - B_R sin\theta_{\tilde q} {\bar q}_R \chi_{1L}) }      
 & & \nonumber \\
& & \frac{1}{q^2-m_{ {\tilde q}_1^2}} (B_L cos \theta_q {\bar \chi}_{1R} q_L
 - B_R sin\theta_{{\tilde q}} {\bar \chi}_{1L} q_R) \nonumber \\
 & & + (B_L sin \theta_q q_L
\chi_{1R} + cos\theta_{\tilde q} {\bar q}_R \chi_{1L}) \frac{1}{q^2-
m _{{\tilde q}_2^2}} \nonumber \\
 & &(B_L sin \theta_q {\bar \chi}_{1R} q_L
 + B_R cos\theta_{\tilde q} {\bar \chi}_{1L} q_R) \} \label{eq:eg 18}
\earr

The above effective interaction can be written as
\begin{equation}
{\it L}_{eff} = {\it L}_{eff}^{LL+RR} + {\it L}_{eff}^{LR}
\end{equation}
The first term involves quarks of the same chirality and is not much effected
by the mixing (provided that it is small). The second term involves quarks 
of opposite chirality and is proportional to the s-quark mixing.
\subsubsection{ The part ${\it L}_{eff}^{LL+RR}$}
Employing a Fierz transformation ${\it L}_{eff}^{LL+RR}$ can be cast in the more
convenient form
 \barr 
{\it L}_{eff}^{LL+RR} =&& (g \sqrt{2})^2  2(-\frac{1}{2})\{ |B_L|^2 
\nonumber  \\
&& (\frac{cos^2 \theta_{\tilde q}}{q^2-m_{ {\tilde q}_1^2}} + 
\frac{sin^2\theta_{\tilde q}}{q^2-m_{ {\tilde q}_2^2}})
 {\bar q}_L \gamma_\lambda q_L
\chi_{1R} \gamma^\lambda \chi_{1R} 
\nonumber  \\
&& + |B_R|^2 (\frac {sin^2 \theta_{\tilde q}}{q^2-m_{ {\tilde q}_1^2}} + 
\frac {cos^2 {\theta_{\tilde q}}} {q^2-m_{{\tilde q}_2^2}} )
 {\bar q}_R \gamma_\lambda q_R
\chi_{1L} \gamma^\lambda \chi_{1L} \} 
 \label{eq:eg.19}
\earr
The factor of 2 comes from the Majorana nature of LSP and the (-1/2) comes
from the Fierz transformation. Equation (\ref{eq:eg.19}) can be
written more compactly as
\barr
{\it L}_{eff} & = &  - \frac {G_F} {\sqrt {2}} 2\{ {\bar q} \gamma_\lambda
(\beta_{0R} +\beta_{3R} \tau_3) (1+ \gamma_5) q
\nonumber \\
 & - & {\bar q} \gamma_\lambda (\beta_{0L} +\beta_{3L} \tau_3)(1-\gamma_5)
 q \}({\bar \chi}_1 \gamma^\lambda \gamma^5 \chi_1 \}
 \label{eq:eg 21}
\earr
with
\barr
\beta_{0R} &=& \Big( \frac {4} {9} \chi^2_{{\tilde u}_R} 
+\frac {1} {9} \chi^2_{{\tilde d}_R}\Big) |C_{11} tan \theta_W|^2\nonumber \\
\beta_{3R} &=& \Big( \frac {4} {9} \chi^2_{{\tilde u}_R} 
-\frac {1} {9} \chi^2_{{\tilde d}_R} \Big) |C_{11} tan \theta_W|^2
\label{eq:eg 22}\\ 
\beta_{0L} &=& | \frac {1} {6} C^R_{11} tan\theta_W
+\frac{1}{2} C^R_{21}|^2 \chi^2_{{\tilde u}_L} + | \frac {1} {6} C^R_{11}
tan\theta_W  - \frac{1}{2} C^R_{21}|^2 \chi^2_{{\tilde d}_L} \nonumber \\
\beta_{3L} &=& | \frac {1} {6} C^R_{11} tan\theta_W +\frac{1}{2} C^R_{21}|^2
\chi^2_{{\tilde u}_L} - | \frac {1} {6} C^R_{11} tan\theta_W 
-\frac{1}{2} C^R_{21}|^2 \chi^2_{{\tilde d}_L} \nonumber 
\earr
with
\beq
\chi^2_{qL} = c^2_{\tilde q} \frac {m_W^2}{m_{{\tilde q}^2_1}-q^2} +
 s^2_{{\tilde q}} \frac {m_W^2}{m_{{\tilde q}^2_2}-q^2}~~,~~
\chi^2_{qR} = s^2_{\tilde q} \frac{m_W^2}{m_{{\tilde q}^2_1}-q^2} +
 c^2_{\tilde q} \frac{m_W^2}{m_{{\tilde q}^2_2}-q^2}
 \label{eq:eg 23}
\eeq
where $c_{\tilde q} =cos\theta_{\tilde q}, \,\ s_{\tilde q}=sin\theta_{\tilde q}$. The above parameters are functions of the four-momentum transfer which
in our case is negligible. \\
 Eq (\ref{eq:eg 21}) it is often written as:
\beq
{\it L}_{eff}  =   - \frac {G_F} {\sqrt {2}} 2\ \left[{\bar u} \gamma_\lambda
(d^0 (u)+ \gamma_5 d (u)) u+ {\bar d} \gamma_\lambda
(d^0 (d)+ \gamma_5 d (d)) d \right ]
 ({\bar \chi}_1 \gamma^\lambda \gamma^5 \chi_1 \}
 \label{eq:eg 21a}
\eeq
where
\beq
d^0(u)=\beta_{0R}+\beta_{3R}-\beta_{0L}-\beta_{3L},d(u)=\beta_{0R}+\beta_{3R}+\beta_{0L}+\beta_{3L}
\label{dofu}
\eeq
\beq
d^0(d)=\beta_{0R}-\beta_{3R}-\beta_{0L}+\beta_{3L},d(u)=\beta_{0R}-\beta_{3R}+\beta_{0L}-\beta_{3L}
\label{dofd}
\eeq
Proceeding as in sec. \ref{sec:Z-exc} we can obtain the
effective Lagrangian at the nucleon level as
\beq
{\it L}_{eff}^{LL+RR} = - \frac {G_F}{\sqrt 2} ({\bar \chi}_1 \gamma^{\lambda}
\gamma^5 \chi_1) J_{\lambda} ({\tilde q})
 \label{eq:eg.24}
\eeq
\beq
J_{\lambda}({\tilde q}) = {\bar N} \gamma_{\lambda} \{f^0_V({\tilde q}) + f^1_V
({\tilde q}) \tau_3 + f^0_A({\tilde q}) \gamma_5 + f^1_A({\tilde q}) \gamma_5
\tau_3) N
  \label{eq:eg.25}
\eeq
with
\barr
 f^0_V = 6(\beta_{0R}-\beta_{0L}) , \qquad f^1_V = 2 g_V
(\beta_{3R}-\beta_{3L})
\nonumber \\
f^0_A = 2g^0_A (\beta_{0R}+\beta_{0L}), \qquad
f^1_A = 2g_A(\beta_{3R}+\beta_{3L})
\label{eq:eg 25a}
\earr
 with $g_v=1.0$ and $g_A=1.25$. The quantity $g^0_A$ depends on the
 quark model for the nucleon. It can be anywhere between 0.12 and 1.00 (see below \ref{sec:nuc-spin}).

We should note that this interaction is more suppressed than the ordinary
weak interaction by the fact that the masses of the s-quarks are usually
larger than that of the gauge boson $Z^0$. In the limit in which the LSP 
is a pure bino ($C_{11} = 1,  C_{21} = 0$) we obtain
\beq
\beta_{0R} = tan^2 \theta_W \Big( \frac{4}{9} \chi^2_{u_R} 
+\frac{1} {9} \chi^2_{{\tilde d}_R}\Big)~,~
\beta_{3R} = tan^2 \theta_W \Big( \frac{4} {9} \chi^2_{u_R} 
-\frac {1} {9} \chi^2_{{\tilde d}_R} \Big) 
\eeq
\beq
\beta_{0L} =  \frac{tan^2 \theta_W} {36} (\chi^2_{{\tilde u}_L} + 
 \chi^2_{{\tilde d}_L}) ~,~
\beta_{3L} =  \frac{tan^2 \theta_W} {36}
(\chi^2_{{\tilde u}_L} -  \chi^2_{{\tilde d}_L})
\label{eq:eg 24}
\eeq
Assuming further that $\chi_{{\tilde u}_R} = \chi_{{\tilde d}_R} 
= \chi_{{\tilde u}_L} = \chi_{{\tilde d}_L}$ we obtain
\beq
 f^1_V ({\tilde q}) / f^0_V({\tilde q}) \simeq + \frac{2}{9}~,~
  f^1_A ({\tilde q})/ f^0_A({\tilde q}) \simeq + \frac{6}{11}
 \label{eq:eg 25}
\eeq
If, on the other hand, the LSP were the photino ($C_{11} = cos\theta_W,
C_{21} = sin \theta_W, C_{31} = C_{41} = 0$) and the s-quarks were
degenerate there would be no coherent contribution ($f^0_V = 0$ if
$\beta_{0L} =\beta_{0R}$).

As we have mentioned in the previous section, to leading order, only the axial current contributes
to the direct detection of the neutralino.
\subsubsection{ The part ${\it L}_{eff}^{LR}$}
From Eq. (\ref{eq:eg 18}) we obtain
\barr
{\it L}_{eff}^{LR}& =& - (g \sqrt{2})^2 sin2\theta_{\tilde q} B_L(q) B_R(q)
\frac{1}{2} [ \frac {1} {q^2-m_{ {\tilde q}_1^2}} - 
\frac{1}{q^2-m_{{\tilde q}_2^2}}]\\
\nonumber
& &({\bar q}_L \chi_{1R} {\bar \chi}_{1L} q_R + 
{\bar q}_R \chi_{1L} {\bar \chi}_{1R} q_L)
\earr
Employing a Fierz transformation we can cast it in the form
\beq
{\it L}_{eff} = - \frac{G_F} {\sqrt{2}} \sum_q \beta(q) \left [({\bar q} q 
{\bar \chi}_1 \chi_1
+ {\bar q}\gamma_5 q {\bar \chi}_1 \gamma_5\chi_1 -({\bar q}\sigma_{\mu\nu} q)
({\bar \chi}_1 \sigma^{\mu\nu}\chi_1)) \right ]
\label{betaq} 
\eeq
where
\beq
\beta(u) = \frac{2}{3} tan\theta_W C_{11} \{ 2sin 2\theta_{\tilde u}
[\frac{1}{6} CR_{11} tan\theta_W + \frac{1}{2} CR_{21}] \Delta_{\tilde u}
\label{betau}
\eeq
\beq
 \beta(d)=sin 2 \theta_{\tilde d}
[\frac{1}{6} CR_{11} tan\theta_W -  \frac{1}{2} C^R_{21}] \Delta_{\tilde d} \}  
\label{betad}
\eeq
Where in the last expressions $u$ indicates quarks with
charge 2/3 and d quarks with charge -1/3. In all cases
\begin{center}
$\Delta_{\tilde u} = \frac{(m^2_{{\tilde u}_1}-m^2_{{\tilde u}_2}) M^2_W}
{(m^2_{{\tilde u}_1}-q^2)(m^2_{{\tilde u}_2}-q^2)}$
\end{center}
and an analogous equation for $\Delta_{\tilde d}$. 

The appearance of scalar terms in s-quark exchange  \cite{ref1} has been first noticed
by Griest \cite{GRIEST}. It has also been noticed there that one should consider
explicitly the effects of quarks other than u and d~\cite{Dree} in going 
from the quark to the nucleon level. We first notice that with the exception 
of $t$ s-quark the ${\tilde q}_L - {\tilde q}_R$ mixing small. Thus
\beq
sin 2\theta_{\tilde u} \Delta {\tilde u} \simeq \frac{2m_u(A + \mu cot\beta)
m^2_W}  {(m^2_{{\tilde u}_L}-q^2) (m^2_{{\tilde u}_R}-q^2)}~,~
sin 2\theta_{\tilde d} \Delta {\tilde d} \simeq \frac{2m_d(A + \mu tan\beta)
m^2_W}  {(m^2_{{\tilde d}_L}-q^2) (m^2_{{\tilde d}_R}-q^2)}
\eeq
In going to the nucleon level and ignoring the negligible pseudoscalar and
tensor components we only need modify the above expressions for all
quarks, with the possible exception of the $t$ quarks, by the substitution $m_q \rightarrow f_q m_N$
(see sec. \ref{sec:nuc-sc}).
 For the t s-quark the mixing is complete,
which implies that the amplitude is independent of the top quark mass.
Hence in the case of the top quark we may  not get an extra enhancement in
going from the quark to the nucleon level. 
In any case this way we get   
\beq
{\it L}_{eff} = \frac{G_F} {\sqrt {2}} [f^0_S ({\tilde q}) {\bar N} N +
 f^1_S ({\tilde q}) {\bar N} \tau_3 N] {\bar \chi}_1 \chi_1
\eeq
with 
\beq
f^0_S ({\tilde q}) = \frac{f_u \beta(u)+f_d \beta(d)}{2}+\sum_{q=s,c,b,t} f_q \beta(q)
\label{beta0}
\eeq
\beq
 f^1_S ({\tilde q}) = \frac{f_u \beta(u)-f_d \beta(d)}{2}
\label{beta1} 
\eeq
(see sec. \ref{sec:nuc-sc} for details).
In the allowed SUSY parameter space considered in this
work this contribution can be neglected in front of the Higgs exchange
contribution. This happens because for quarks other than t the s-quark
mixing is small. For the t-quark, as it has already been mentioned, we
have large mixing, but we do not get the advantage of the mass enhancement.
 
\subsection{The Intermediate Higgs Contribution}
\label{sec:Higgs-exc}
\begin{figure}
\begin{center}
\includegraphics[height=8cm]{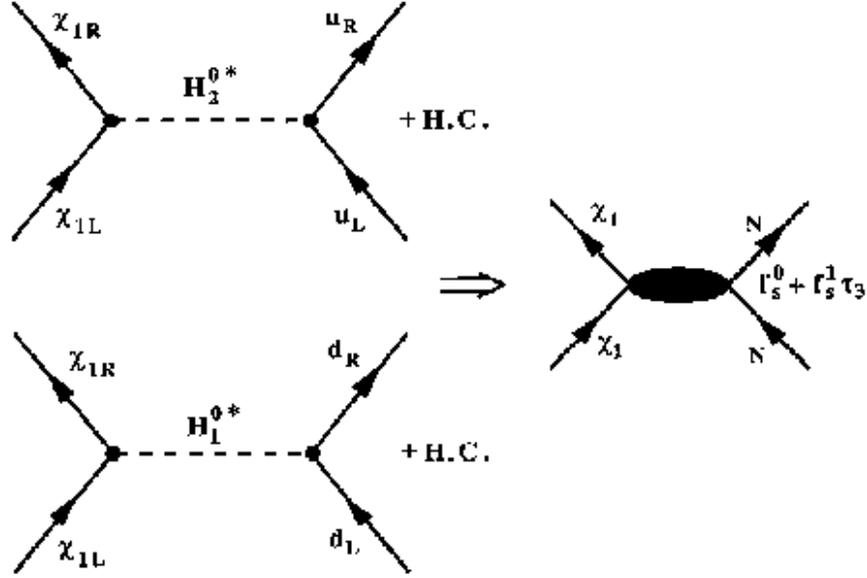}
\caption{ The LSP-quark interaction mediated by Higgs exchange.
 \label{LSPH} }
\end{center}
\end{figure}

The coherent scattering can be mediated via the intermediate Higgs
particles which survive as physical particles (see Fig. \ref{LSPH}).
The relevant interaction can arise out of the
Higgs-Higgsino-gaugino interaction which takes the form
\barr
 {\it L}_{H \chi \chi} &=& \frac {g}{\sqrt 2} \Big({\bar{\tilde W}}^3_R
 {\tilde H}_{2L} H^{0*}_2 - {\bar{\tilde W}}^3_R {\tilde H}_{1L} H^{0*}_1 
\nonumber \\
   &-&tan \theta_w ({\bar{\tilde B}}_R
  {\tilde H}_{2L} H^{0*}_2 - {\bar{\tilde B}}_R {\tilde H}_{1L} H^{0*}_1)
 \Big) + H.C.
 \label{eq:eg 26}
\earr
Proceeding as above we can express ${\tilde W}$ an ${\tilde B}$ in terms
of the appropriate eigenstates and retain the LSP to obtain
\barr
{\it L} &=& \frac {g}{\sqrt 2} \Big((C^R_{21} -tan\theta_w C^R_{11})
C_{41}{\bar \chi}_{1R} \chi_{1L} H^{o*}_2 \nonumber \\
  &-&(C^R_{21} -tan \theta_w C^R_{11}) 
C_{31}{\bar \chi}_{1R} \chi_{1L} H^{o*}_1 \Big) + H.C.
\label{eq:eg 27}
\earr

We can now proceed further and express the fields 
${H^0_1}^*$, ${H^0_2}^*$ in terms of the physical fields $h$, $H$ and
$A.$ The term which contains $A$ will be neglected, since it yields only
a pseudoscalar coupling which does not lead to coherence.

Thus we can write
\beq
{\cal L}_{eff} = - \frac{G_F}{\sqrt{2}}{\bar \chi} \chi \,
{\bar N} [ f^0_s (H) + f^1_s (H) \tau_3 ] N
\label{2.1.1}
\eeq
where
\beq
f^0_S (H)  = \frac{1}{2} (g_u + g_d) + g_s + g_c + g_b + g_t
\label{2.1.2}
\eeq
\beq
f^1_S (H)  = \frac{1}{2} (g_u - g_d)
\label{2.1.3}
\eeq
with
\beq
g_{a_i}  = \big[ g_1(h) \frac{cos \alpha}{sin \beta}
+ g_2(H) \frac{sin \alpha}{sin \beta} \big] \frac{m_{a_i}}{m_N},
\quad a_i = u,c,t
\label{2.1.4}
\eeq
\beq
g_{\kappa_i}  = \big[- g_1(h) \frac{sin \alpha}{cos \beta}
+ g_2(H) \frac{cos \alpha}{cos \beta} \big] 
\frac{m_{\kappa_i}}{m_N},
\quad \kappa_i = d,s,b
\label{2.1.5}
\eeq
\beq
g_{1}(h)  = 4 (C^R_{11 } tan \theta_W - C^R_{21}) (C_{41 } cos \alpha + 
            C_{31} sin \alpha) \frac{m_N m_W}{m^2_h -q^2}
\label{2.1.6}
\eeq
\beq
g_{2}(H)  = 4 (C^R_{11 } tan \theta_W - C^R_{21}) (C_{41 } sin \alpha - 
            C_{31}cos \alpha) \frac{m_N m_W}{m^2_H -q^2}
\label{2.1.7}
\eeq
where $m_N$ is the nucleon mass, and
the parameters $m_h$, $m_H$ and $\alpha$ depend on the SUSY parameter
space (see Table 1). 
\section{Going from the Quark to the Nucleon Level} 
\label{sec:nuc}
As we have already mentioned, one has to be a bit more 
careful in handling quarks other than $u$ and $d$.
\subsection{The scalar interaction}
\label{sec:nuc-sc}
 As we have seen the scalar couplings of the LSP to the quarks
are proportional to their mass~\cite{Dree}.
One encounters in the nucleon not only
sea quarks ($u {\bar u}, d {\bar d}$ and $s {\bar s}$) but the heavier
quarks also due to QCD effects ~\cite{Dree00}. 
This way one obtains the scalar Higgs-nucleon
coupling by using  effective parameters $f_q$ defined as follows:
\beq
\Big<N| m_q \bar{q}q|N \Big> = f_q m_N
\label{fofq}
\eeq
where $m_N$ is the nucleon mass. 
 The parameters $f_q,~q=u,d,s$ can be obtained by chiral
symmetry breaking 
terms in relation to phase shift and dispersion analysis.
The isoscalar component can be obtained  by considering the following
quantities
:
\begin{enumerate}
\item The phenomenologically determined mass ratios:
\beq
\frac{m_u}{m_d}=0.553 \pm 0.043~~,~~\frac{m_s}{m_d}=18.9 \pm 0.08
\label{mumd}
\eeq
\item The quantities :
\beq
z=\frac{B_u-B_s}{B_d-B_s} \approx 1.49~~,~~y=\frac{2 B_s}{B_d+B_u}
\label{zbubd}
\eeq
One then finds that:
\begin{equation}
\frac{B_u}{B_d}=\frac{2z-(z-1)y}{2+(z-1)y} \mbox{ for protons}~,
~\frac{B_u}{B_d}=\frac{2+(z-1)y}{2z-(z-1)y} \mbox{ for neutrons}
\label{proton1}
\end{equation}
with $B_q=<N|\bar{q}q|N>$
\item  The pion-nucleon sigma-term, $\sigma_{\pi N}$: \\
this term is obtained from
the isospin even $\pi$-N scattering amplitude with vanishing external momenta.
It is defined by the scalar quark density operator averaged over the nucleon
or equivalently by $\sigma_{\pi N}(t=0)$,
the scalar form factor of the nucleon at zero momentum transfer squared.
The value of the sigma term is deduced from the analysis of two
quantities: $\sigma_{\pi N}(t=2M_{\pi}^2)$ the scalar form factor at the
Cheng-Dashen point, which is experimentally accessible, and the difference
$\Delta_\sigma =
\sigma_{\pi N}(2M_{\pi}^2) - \sigma_{\pi N}(0) = 15.2 \pm 0.4$
MeV~\cite{SAINIO91},\cite{GASSER82} as induced by explicit chiral symmetry breaking.
Experimentally, after efforts of many years, the value of the sigma-tern
is still quite uncertain \cite{GASSER82}. The canonical value of the $\pi N$
sigma term with
\begin{equation}\label{sigmaold}
\sigma_{\pi N} = {m_u + m_d \over 2} (B_u + B_d)=(45\pm 8) ~MeV
\end{equation}
is deduced from an earlier analyses with
$\sigma_{\pi N}(t=2M_{\pi}^2) = 60 \pm 8$ \cite{SAINIO91}.
During the last few years analyses of also more recent pion-nucleon
scattering data lead to an increase in the value of scalar form factor
at the Cheng-Dashen point $\sigma_{\pi N} (M_{\pi}^2)$ with
$88 \pm 15$ MeV~\cite{Kaufmann:dd},
$71 \pm 9$ MeV~\cite{Olsson:1999jt}, $79 \pm 7$ MeV~\cite{Pavan:2001wz}
and $(80-90)$ MeV \cite{Olsson:2002}. Thus the recent analyses suggest
a value for the pion-nucleon sigma term of about
\begin{equation}\label{sigmanew}
\sigma_{\pi N} = {m_u + m_d \over 2} (B_u + B_d)=(56 - 75) ~MeV
\end{equation}.
\item Theoretical analysis of the $\sigma_{\pi N}$ term:\\
 In the context of chiral
perturbation one can show that:
\begin{equation}\label{yequation}
\sigma_{\pi N}= {\sigma_0 \over 1-y}~,~\sigma_0=(35\pm 5)  ~MeV
\end{equation} 
Eqs. (\ref{sigmaold}) and (\ref{sigmanew}) together with Eq. (\ref{yequation}) will provide
the range of variation in the parameter y.
Taking:
\begin{equation}
m_u = 5.1~MeV ~~,m_d=9.3~MeV
\end{equation}
together with y will in turn provide by Eq. \ref{proton1} the range of variation of the ratio
 $B_u/B_d$.
The uncertainties in Eqs. (\ref{sigmaold},\ref{sigmanew},\ref{yequation}) provide a wide range
in which the parameter y can vary. 
In other words the experimental and theoretical uncertainties permit us, we will exploit the possible
consequences of variation in y to SUSY dark matter detection.
For $\sigma_{\pi N}$ we choose $45,~55,~65$ and $75$ MeV to reflect the range of values
set by Eqs. (\ref{sigmaold}) and (\ref{sigmanew}). Thus from Eq. (\ref{yequation}) we extract
the corresponding y parameters with $0.22\pm 0.11,~0.36\pm 0.09,~046\pm 0.08$ and
$0.53\pm 0.07$, respectively. Then we will combine these values with Eq. \ref{proton1} to get the desired
values of $f_q$ given below.
\end{enumerate}
 From the above analysis we get in the case of the proton:
\begin{equation}
f^p_d=\frac{\Sigma_{\pi N}}{0.756~m_p}[1+ \frac{2z-(z-1)y}{2+(z-1)y}]^{-1}
\label{fpd}
\end{equation}
\begin{equation}
f^p_u=0.553~f^p_d~[\frac{2z-(z-1)y}{2+(z-1)y}]~,~f^p_s=\frac{\Sigma_{\pi N}}{0.756~m_p}~\frac{19}{2}~y
\label{fpu}
\end{equation}
\begin{equation}
f^p_s=\frac{\Sigma_{\pi N}}{0.756~m_p}~\frac{19}{2}~y
\label{fps}
\end{equation}
In the case of the neutron our expressions are analogous, the ratio $B_u/B_d$
getting the inverse value.\\
For the heavy quarks, to leading order via quark loops and gluon exchange with the
nucleon, one finds:
\begin{center}
\quad $f_Q= 2/27(1-\quad \sum_q f_q)$   
\end{center}
 There is a correction to the above parameters coming from loops
involving s-quarks 
\cite {Dree00}.
The leading contribution can be absorbed into
the definition, if the functions $g_1(h)$ and $g_2(H)$ as follows :
\begin{center}
\quad $g_1(h) \rightarrow g_1(h)[1+\frac{1}{8}(2 \frac{m^2_Q}{m^2_W}
-\frac{sin(\alpha+\beta)}{cos^2\theta_W}\frac{sin\beta}{cos\alpha})]$   
\end{center}
\begin{center}
\quad $g_2(H) \rightarrow g_1(h)[1+\frac{1}{8}(2 \frac{m^2_Q}{m^2_W}
+\frac{cos(\alpha+\beta)}{cos^2\theta_W}\frac{sin\beta}{sin\alpha})]$   
\end{center}
for $Q=c$ and $t$ For the b-quark we get:
\begin{center}
\quad $g_1(h) \rightarrow g_1(h)[1+\frac{1}{8}(2 \frac{m^2_b}{m^2_W}
-\frac{sin(\alpha+\beta)}{cos^2\theta_W}\frac{cos\beta}{cos\alpha})]$   
\end{center}
\begin{center}
\quad $g_2(H) \rightarrow g_1(h)[1+\frac{1}{8}(2 \frac{m^2_b}{m^2_W}
-\frac{cos(\alpha+\beta)}{cos^2\theta_W}\frac{cos\beta}{sin\alpha})]$   
\end{center}
In addition to the above effects one has to consider QCD effects. 
These effects renormalize the contribution of the quark loops as follows \cite {Dree00}:
\begin{center}
\quad $f_{QCD}(q)=\frac{1}{4}\frac{\beta(\alpha_s)}
{1+\gamma_m(\alpha_s)}$
\end{center}
with
\begin{center}
$\beta(\alpha_s)=\frac{\alpha_s}{3 \pi}[1+\frac{19}{4}{\alpha_s}{\pi}]$ ,
$\gamma_m(\alpha_s)=2\frac{\alpha_s}{\pi}$
\end{center}
Thus
\begin{center}
\quad $f_{QCD}(q)=1+\frac{11}{4}\frac{\alpha_s}{\pi}$
\end{center}
The QCD correction associated with the s-quark loops is:
\begin{center}
\quad $f_{QCD}(\tilde{q})=1+\frac{25}{6}\frac{\alpha_s}{\pi}$
\end{center}
 The above corrections depend on Q since $\alpha_s$ must be evaluated
at the scale of $m_Q$. \\
 It convenient to introduce the factor$f_{QCD}(\tilde{q})/f_{QCD}(q)$
into the factors $g_1(h)$ and $g_2(H)$ and the factor of $f_{QCD}(q)$
into the the quantities $f_Q$. If, however, one restricts oneself to
the large $tan\beta$ regime, the corrections due to the s-quark loops
is independent of the parameters $\alpha$ and $\beta$ and significant
only for the t-quark.

For a more detailed discussion we refer the reader to 
Refs.~\cite{Dree,Dree00}. We thus obtain the results presented in Table \ref{table.fq}.
\begin{table}[t]
\caption{ The parameters $f^p_q$ and $f^p_Q$ (upper part) as well as
 $f^n_q$ and $f^n_Q$ (lower part) for the twelve cases discussed in
the text.
}
\label{table.fq}
\begin{center}
\begin{tabular}{rrrrrrrrrrrrrr}
 &   &  &  &  &   &  &  &  &  &  &  &  &\\
$\#$ & $f^p_d$ & $f^p_u$ & $f^p_s$ & $f^p_c$ & $f^p_b $ & $f^p_t$& &$f^n_d$ & $f^n_u$ & $f^n_s$ & $f^n_c$ & $f^n_b $ & $f^n_t$\\
\hline
 &   &  &  &  &   &  &  &  &  &  &  &  &\\
 1& 0.026& 0.021& 0.067& 0.098& 0.104& 0.161& &0.037& 0.014& 0.066& 0.098& 0.104& 0.161\\
 2& 0.027& 0.020& 0.133& 0.087& 0.092& 0.144& &0.037& 0.015& 0.133& 0.086& 0.092& 0.143\\
 3& 0.028& 0.020& 0.199& 0.075& 0.080& 0.126& &0.036& 0.015& 0.199& 0.075& 0.080& 0.126\\
 4& 0.033& 0.025& 0.199& 0.078& 0.083& 0.132& &0.044& 0.018& 0.199& 0.077& 0.083& 0.122\\
 5& 0.034& 0.024& 0.265& 0.068& 0.072& 0.117& &0.044& 0.019& 0.265& 0.067& 0.172& 0.117\\
 6& 0.031& 0.025& 0.332& 0.057& 0.061& 0.106& &0.043& 0.017& 0.332& 0.057& 0.062& 0.102\\
 7& 0.040& 0.028& 0.331& 0.061& 0.065& 0.109& &0.051& 0.022& 0.331& 0.060& 0.065& 0.109\\
 8& 0.041& 0.028& 0.400& 0.051& 0.055& 0.095& &0.051& 0.023& 0.400& 0.050& 0.055& 0.095\\
 9& 0.047& 0.028& 0.470& 0.041& 0.047& 0.081& &0.051& 0.023& 0.400& 0.050& 0.055& 0.095\\
10& 0.047& 0.027& 0.462& 0.045& 0.050& 0.090& &0.050& 0.023& 0.470& 0.040& 0.044& 0.060\\
11& 0.048& 0.032& 0.532& 0.036& 0.040& 0.076& &0.058& 0.027& 0.532& 0.035& 0.040& 0.076\\
12& 0.049& 0.032& 0.603& 0.026& 0.030& 0.063& &0.057& 0.027& 0.603& 0.026& 0.030& 0.063\\
\end{tabular}
\end{center}
\end{table}

We notice that there exist differences between the proton and neutron
components. These, however, cannot be taken as the sole contribution to
 isovector contribution, since all quantities were derived with isoscalar
operators. So the isovector contribution will be discussed elsewhere.
Here we will limit ourselves to the isoscalar component $f_q=(f^p_q+f^n_q)/2$

\subsection{The axial current contribution}
\label{sec:nuc-spin}
The amplitudes $a_p=f^0_A+f^1_A$ and $a_n=f^0_A-f^1_A$ are defined by \cite{JELLIS}:
\beq
 a_N=\sum_{q=u,d,s}d_q \Delta q_N
 \label{amplitudea}
 \eeq
 \beq
 2s_{\mu}\Delta q_N=\left<N|\bar{q}\gamma_{\mu}\gamma_5 q|N\right >
 \label{deltaq}
 \eeq
  where $s_{\mu}$ is the nucleon spin and $d_q$ the relevant spin amplitudes at the quark level obtained in a
  given SUSY model.\\
  The isoscalar and the isovector axial current
couplings at the nucleon level, $ f^0_A$, $f^1_A$, are obtained from the corresponding ones given by the SUSY
 models at the quark level, $ f^0_A(q)$, $f^1_A(q)$, via renormalization
coefficients $g^0_A$, $g_A^1$, i.e.
$ f^0_A=g_A^0 f^0_A(q),f^1_A=g_A^1 f^1_A(q).$
The  renormalization coefficients are given terms of $\Delta q$ defined above \cite{JELLIS},
via the relations
$$g_A^0=\Delta u+\Delta d+\Delta s=0.77-0.49-0.15=0.13~,~g_A^1=\Delta u-\Delta d=1.26$$
We see that, barring very unusual circumstances at the quark level, the isoscalar contribution is
negligible. It is for this reason that one might prefer to work in the isospin basis.
\section{The nucleon cross sections}
\label{sec:nuc-sigma}
With the above ingredients we are in a position to evaluate the nucleon cross sections. 
\begin{itemize}
\item The scalar cross section. As we have mentioned this is primarily due to the Higgs exchange.
\beq
\sigma^S_{p,\chi^0}=\sigma_0 |{f^0_S}+{f^1_S}|^2~,~\sigma^S_{n,\chi^0}=\sigma_0 |{f^0_S}-{f^1_S}|^2
\label{sigmasc}
\eeq
with
\beq
\sigma_0 = \frac{1}{2\pi} (G_F m_p)^2 = 0.77 \times 10^{-38}
cm^2 = 0.77 \times 10^{-2} pb
\label{sigma0}
\eeq
Since, however, the process is dominated by quarks other than $u$ and $d$, the isovector contribution is negligible.
So we can talk about the nucleon cross section.
\item The proton spin cross section is given by:
 \beq
\sigma^{spin}_{p,\chi^0}=3 \sigma_0 |{f^0_A}+{f^1_A}|^2=3 \sigma_0|a_p|^2
\label{eq:eg 52}
\eeq
\end{itemize}
\section{The allowed SUSY Parameter Space}
\label{sec:parameter}
 It is clear from the above discussion that the nucleon cross section depends:
\begin{itemize}
\item The the quark structure of the nucleon\\
The allowed range of the parameters $f_q$ may induce variations in the nucleon cross section as large as an order
of magnitude.
\item The parameters of supersymmetry.\\
This is the most crucial input. One starts with a set of parameters at the GUT scale and predicts the low energy
observable via the renormalization group equations (RGE). Conversely starting from the low energy phenomenology
one can constrain the input parameters at the GUT scale. 
\end{itemize}
 The parameter space is the most crucial. In SUSY models derived from minimal SUGRA
the allowed parameter space is characterized at the GUT scale  by five 
parameters:
\begin{itemize}
\item two universal mass parameters, one for the scalars, $m_0$, and one for the
fermions, $m_{1/2}$.
\item $tan\beta $.
\item The trilinear coupling  $ A_0 $ (or  $ m^{pole}_t $)  and 
\item The sign of $\mu $ in the Higgs self-coupling  $\mu H_1~H2$.
\end{itemize}
The experimental constraints are
\begin{enumerate}
\item The LSP relic abundance (including co-annihilations):
$$0.09 \leq\Omega_{LSP}h^2\leq0.22 \mbox{ (previous)},~0.09 \leq\Omega_{LSP}h^2\leq0.124 \mbox{ (WMAP)} $$
\item the $b\rightarrow s \gamma$ constraint (CLEO, BELLE) 
$$2\times 10^{-4}\le BR \le 4.1\times 10^{-4}$$
\item The Higgs mass:$~~~ \geq 114.1~ GeV$. This applies on the standard model Higgs. So
For SUSY one must correct for factor $\sin^2{(\alpha-\beta)}$
where $\alpha$ is the Higgs mixing angle. So this imposes limits on $\tan{\beta}$
\item Limits on $g_s-2$ $(e^-,e^+)$ experiments (E821 at BNL)
 $$a_{\mu}=(g_{\mu}-2)/2=(33.7\pm11.2)\times10^{-10}$$
yields ($2 \sigma$ level):
     $$11.3 \times 10^{-10} \leq \delta a_{\mu}(SUGRA)
 \leq 56.1 \times 10^{-10}$$
\item The fermion masses:\\
$m_t(pole)=175~GeV~,~m_b(m_b)=4.25~GeV \Rightarrow$\\
 $m_b(m_Z)=2.888~GeV,~m_{\tau}(M_Z)=1.7463~GeV$
\end{enumerate}
 We are not going to elaborate further on this interesting aspect, since it will be covered by another contribution to these
proceedings by A. Lahanas.
\section{Rates}
\label{sec:rates}
The differential non directional  rate can be written as
\begin{equation}
dR_{undir} = \frac{\rho (0)}{m_{\chi}} \frac{m}{A m_N}
 d\sigma (u,\upsilon) | {\boldmath \upsilon}|
\label{2.18}
\end{equation}
where A is the nuclear mass number, 
$\rho (0) \approx 0.3 GeV/cm^3$ is the LSP density in our vicinity,
 m is the detector mass, 
 $m_{\chi}$ is the LSP mass and $d\sigma(u,\upsilon )$ is the differential cross section.
 
 The directional differential rate, i.e. that obtained, if nuclei recoiling in the direction $\hat{e}$ are 
observed, is given by \cite{JDVSPIN04}:
\beq
dR_{dir} = \frac{\rho (0)}{m_{\chi}} \frac{m}{A m_N}
|\upsilon| \hat{\upsilon}.\hat{e} ~\Theta(\hat{\upsilon}.\hat{e})
 ~\frac{1}{2 \pi}~
d\sigma (u,\upsilon\
\nonumber \delta(\frac{\sqrt{u}}{\mu_r \upsilon
\sqrt{2}}-\hat{\upsilon}.\hat{e})
 \label{2.20}
\eeq
where $\Theta(x)$ is the Heaviside function.

The differential cross section is given by:
\beq
d\sigma (u,\upsilon)== \frac{du}{2 (\mu _r b\upsilon )^2}
 [(\bar{\Sigma} _{S}F(u)^2
                       +\bar{\Sigma} _{spin} F_{11}(u)]
\label{2.9}
\end{equation}
where $ u$ the energy transfer $Q$ in dimensionless units given by
\begin{equation}
 u=\frac{Q}{Q_0}~~,~~Q_{0}=[m_pAb]^{-2}=40A^{-4/3}~MeV
\label{defineu}
\end{equation}
 with  $b$ is the nuclear (harmonic oscillator) size parameter. $F(u)$ is the
nuclear form factor and $F_{11}(u)$ is the spin response function associated with
the isovector channel.

The scalar
cross section is given by:
\begin{equation}
\bar{\Sigma} _S  =  (\frac{\mu_r}{\mu_r(p)})^2
                           \sigma^{S}_{p,\chi^0} A^2
 \left [\frac{1+\frac{f^1_S}{f^0_S}\frac{2Z-A}{A}}{1+\frac{f^1_S}{f^0_S}}\right]^2
\approx  \sigma^{S}_{N,\chi^0} (\frac{\mu_r}{\mu_r (p)})^2 A^2
\label{2.10}
\end{equation}
(since the heavy quarks dominate the isovector contribution is
negligible). $\sigma^S_{N,\chi^0}$ is the LSP-nucleon scalar cross section.
The spin Cross section is given by:
\begin{equation}
\bar{\Sigma} _{spin}  =  (\frac{\mu_r}{\mu_r(p)})^2
                           \sigma^{spin}_{p,\chi^0}~\zeta_{spin},
\zeta_{spin}= \frac{1}{3(1+\frac{f^0_A}{f^1_A})^2}S(u)
\label{2.10a}
\end{equation}
\begin{equation}
S(u)\approx S(0)=[(\frac{f^0_A}{f^1_A} \Omega_0(0))^2
  +  2\frac{f^0_A}{ f^1_A} \Omega_0(0) \Omega_1(0)+  \Omega_1(0))^2  \, ]
\label{s(u)}
 \end{equation}
 The couplings $f^1_A$ ($f^0_A$) and the nuclear matrix elements $\Omega_1(0)$ ($\Omega_0(0)$) associated
 with the isovector (isoscalar) components are normalized so that, in the case
 of the proton at $u=0$, they yield $\zeta_{spin}=1$.

 With these definitions in the proton neutron representation we get:
 \beq
 \zeta_{spin}= \frac{1}{3}S^{'}(0)
 \label{zeta3}
 \eeq
 \beq
 S^{'}(0)=\left[(\frac{a_n}{a_p}\Omega_n(0))^2+2 \frac{a_n}{a_p}\Omega_n(0) \Omega_p(0)+\Omega^2_p(0)\right]
 \label{Spn}
 \eeq
 where $\Omega_p(0)$ and $\Omega_n(0)$ are the proton and neutron components of the static spin nuclear matrix elements. In extracting limits on the nucleon cross sections from the data we will find it convenient to
 write:
 \begin{equation}
                          \sigma^{spin}_{p,\chi^0}~\zeta_{spin} =\frac{\Omega^2_p(0)}{3}|\sqrt{\sigma_p}+\frac{\Omega_n}{\Omega_p} \sqrt{\sigma_n}
 e^{i \delta}|^2
\label{2.10ab}
\end{equation}
 In Eq. (\ref{2.10ab}) $\delta$ the relative
phase between the two amplitudes $a_p$ and $a_n$.
 The static spin matrix elements are obtained in the context of a given nuclear model. Some such matrix elements of interest to the planned experiments are given in table \ref{table.spin}.
 The shown results
are obtained from Divari \cite{DIVA00}, Ressel {\it et al} (*) \cite{Ress},
 the Finish group (**) \cite {SUHONEN03} and the Ioannina team (+) \cite{ref1}, \cite{KVprd}.

\begin{table}[t]
\caption{
 The static spin matrix elements for various nuclei. For $^3$He see Moulin, Mayet and Santos
\cite{Santos},\cite{SANTOS04}. For the other
light nuclei the calculations are from DIVARI \cite{DIVA00}.
 For  $^{73}$Ge and $^{127}$I the results presented  are from Ressel {\it et al}
\cite{Ress} (*) and the Finish group {\it et al} \cite {SUHONEN03}
 (**).
 For $^{207}$Pb they were obtained by the Ioannina team (+).
\cite{ref1}, \cite{KVprd}.
 \label{table.spin} }
\begin{center}
\begin{tabular}{lrrrrrrrr}
\hline\hline
 &   &  &  &  &   &  & &\\
 &$^3$ He& $^{19}$F & $^{29}$Si & $^{23}$Na  & $^{73}$Ge & $^{127}$I$^*$ & $ ^{127}$I$^{**}$ & $^{207}$Pb$^+$\\
\hline
    &   &  &  &  &    \\
$\Omega_{0}(0)$ &1.244     & 1.616   & 0.455  & 0.691  &1.075 & 1.815  &1.220  & 0.552\\
$\Omega_{1}(0)$&-1.527     & 1.675  & -0.461  & 0.588 &-1.003 & 1.105  &1.230  & -0.480\\
$\Omega_{p}(0)$ &-0.141    & 1.646  & -0.003  & 0.640  &0.036 &1.460   &1.225  & 0.036\\
$\Omega_{n}(0)$ &1.386     & -0.030   & 0.459  & 0.051  &1.040 & 0.355  &-0.005 & 0.516\\
$\mu_{th} $& &2.91   &-0.50  & 2.22  &    & &\\
$\mu_{exp}$& &2.62   &-0.56  & 2.22  &    & &\\
$\frac{\mu_{th}(spin)}{ \mu_{exp}}$& &0.91   &0.99  & 0.57  &    &  &\\
\hline
\hline
\end{tabular}
\end{center}
\end{table}
The spin ME are defined as follows:
\beq
\Omega_p(0)=\sqrt{\frac{J+1}{J}}\prec J~J| \sigma_z(p)|J~J\succ ~~,~~
\Omega_n(0)=\sqrt{\frac{J+1}{J}}\prec J~J| \sigma_z(n)|J~J\succ
\label{Omegapn}
\eeq
where $J$ is the total angular momentum of the nucleus and $\sigma_z=2 S_z$. The spin operator is defined by $S_z(p)=\sum_{i=1}^{Z} S_z(i)$, i.e. a sum over all protons in the nucleus,  and
$S_z(n)=\sum_{i=1}^{N}S_z(i)$, i.e. a sum over all neutrons. Furthermore
\beq
\Omega_0(0)=\Omega_p(0)+\Omega_n(0)~~,~~
\Omega_1(0)=\Omega_p(0)-\Omega_n(0)
\label{Omegaiso}
\eeq
\section{Expressions for the Rates}
To obtain the total rates one must fold with LSP velocity distribution and integrate  the
above expressions  over the
energy transfer from $Q_{min}$ determined by the detector energy cutoff to $Q_{max}$
determined by the maximum LSP velocity (escape velocity, put in by hand in the
Maxwellian distribution), i.e. $\upsilon_{esc}=2.84~\upsilon_0$ with  $\upsilon_0$
the velocity of the sun around the center of the galaxy($229~Km/s$).

For a given velocity distribution f(\mbox{\boldmath $\upsilon$}$^{\prime}$),
 with respect to the center of the galaxy,
one can find the velocity distribution in the Lab
f(\mbox{\boldmath $\upsilon$},\mbox{\boldmath $\upsilon$}$_E$)
by writing 
\mbox{\boldmath $\upsilon$}$^{'}$=
          \mbox{\boldmath $\upsilon$}$ \, + \,$ \mbox{\boldmath $\upsilon$}$_E \, ,$
\mbox{\boldmath $\upsilon$}$_E$=\mbox{\boldmath $\upsilon$}$_0$+
 \mbox{\boldmath $\upsilon$}$_1$, with
\mbox{\boldmath $\upsilon$} $_1 \,$ the  Earth's velocity
 around the sun.
  
It is convenient to choose a coordinate system so  that
 $\hat{x}$  is radially out in the plane of the galaxy,
 $\hat{z}$ in the sun's direction of motion and 
 $\hat{y}=\hat{x}\times\hat{z}$.

\begin{figure}
\begin{center}
\includegraphics[height=.3\textheight]{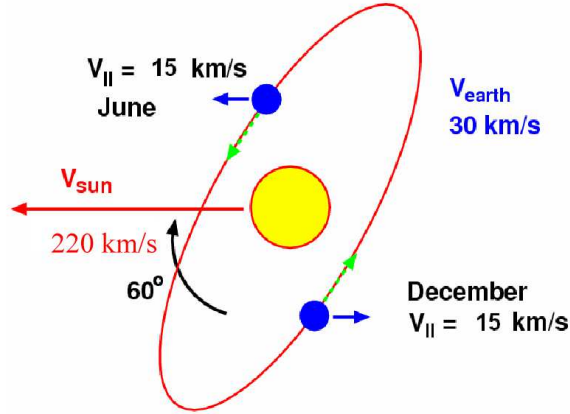}
\caption{The galactic plane is perpendicular to the paper containing the sun's velocity. The normal to the two planes form an angle $\gamma^{'}=\pi/2-\gamma\approx \pi/6$.
 The modulation is affected by the projection of the Earth's velocity along the sun's velocity. Thus the velocity of the detector relative to the center of the galaxy is $220+15=235$ km/s around June 3nd (when the maximum of the event rate is expected) and $220-15=205$ km/s around December 3 (minimum of the event rate).  \label{fig:velocity} }
\end{center}
\end{figure}

Since the axis of the ecliptic  
lies very close to the $x,y$ plane ($\omega=186.3^0$) only  the angle
 $\gamma=29.8^0$ (see Fig. \ref{fig:velocity})
becomes relevant.
Thus the velocity of the earth around the
sun is given by 
\begin{equation}
\mbox{\boldmath $\upsilon$}_E  = \mbox{\boldmath $\upsilon$}_0 \hat{z} +
                                  \mbox{\boldmath $\upsilon$}_1  
(\, sin{\alpha} \, {\bf \hat x}
-cos {\alpha} \, cos{\gamma} \, {\bf \hat y}
+ cos {\alpha} \, sin{\gamma} \, {\bf \hat z} \,)
\label{3.6}  
\end{equation}
where $\alpha$ is  phase of the earth's orbital motion.
The LSP velocity distribution f(\mbox{\boldmath $\upsilon$}$^{\prime}$) is not known. Many velocity distributions
are employed. In the present work we will adopt the standard practice and assume it to be Gaussian:
\beq
f(\mbox{\boldmath $\upsilon$}^{\prime})=\frac{1}{(\sqrt{\pi}\mbox{\boldmath $\upsilon_0$})^3}
e^{-(\mbox{\boldmath $\upsilon$}^{\prime}/\mbox{\boldmath $\upsilon_0$})^2}
\label{fv}
\eeq

 Since \mbox{\boldmath $\upsilon$}$_1 \ll $\mbox{\boldmath $\upsilon$}$_0$ we will ignore, for the moment, the motion of the Earth. Then the total (non directional) rate
is given by
\begin{equation}
R =  \bar{R}\, t(a,Q_{min}) \, \label{3.55f}
\end{equation}
$$ \bar{R}=\frac{\rho (0)}{m_{\chi^0}} \frac{m}{Am_p}~
              (\frac{\mu_r}{\mu_r(p)})^2~ \sqrt{\langle
v^2 \rangle } [\sigma_{p,\chi^0}^{S}~A^2+
 \sigma _{p,\chi^0}^{spin}~\zeta_{spin}]$$
 The SUSY parameters have been absorbed in $\bar{R}$. The
 parameter $t$ takes care of the nuclear form factor and the
 folding with LSP velocity distribution \cite{Verg00,Verg01,JDVSPIN04,JDV06}. It depends on
$Q_{min}$, i.e.  the  energy transfer cutoff imposed by the
detector and $a=[\mu_r b \upsilon _0 \sqrt 2 ]^{-1}$.\\
  In the present work  we find it convenient to re-write it as:
\begin{equation}
R= \bar{K} \left[c_{coh}(A,\mu_r(A)) \sigma_{p,\chi^0}^{S}+
c_{spin}(A,\mu_r(A))\sigma _{p,\chi^0}^{spin}~\zeta_{spin} \right]
\label{snew}
\end{equation}
 where
\beq
\bar{K}=\frac{\rho (0)}{100\mbox{ GeV}} \frac{m}{m_p}~
              \sqrt{\langle v^2 \rangle }\simeq 160~10^{-4}~(pb)^{-1} y^{-1}\frac{\rho(0)}{0.3GeVcm^{-3}}
\frac{m}{1Kg}\frac{ \sqrt{\langle
v^2 \rangle }}{280kms^{-1}}
\label{Kconst}
\eeq
and
\begin{equation}
c_{coh}(A, \mu_r(A))=\frac{100\mbox{ GeV}}{m_{\chi^0}}\left[ \frac{\mu_r(A)}{\mu_r(p)} \right]^2 A~t_{coh}(A)
\label{ctm}
\end{equation}
\begin{equation}
c_{spin}(A, \mu_r(A))=\frac{100GeV}{m_{\chi^0}}\left[ \frac{\mu_r(A)}{\mu_r(p)} \right]^2 \frac{t_{spin}(A)}{A}
\label{ctm1}
\end{equation}
The parameters $c_{coh}(A,\mu_r(A))$, $c_{spin}(A,\mu_r(A))$, which give the relative merit
 for the coherent and the spin contributions in the case of a nuclear
target compared to those of the proton,  have already been  tabulated \cite{JDV06}
 for energy cutoff $Q_{min}=0,~10$ keV.\\
Via  Eq. (\ref{snew}) we can  extract the nucleon cross section from
 the data (see below).
 \\Neglecting the isoscalar contribution and using
$\Omega^2_1=1.22$ and $\Omega^2_1=2.8$ for $^{127}$I and $^{19}$F respectively
the extracted nucleon cross sections satisfy:
\begin{equation}
\frac{\sigma^{spin}_{p,\chi^0}}{\sigma^{S}_{p,\chi^0}} =
 \left[\frac{c_{coh}(A,\mu_r(A))}{c_{spin}(A,\mu_r(A))}\right]
\frac{3 }{\Omega^2_1} \Rightarrow
\approx  \times 10^{4}~(A=127)~,~ \approx \times 10^2~(A=19)
\label{ratior2}
\end{equation}
It is for this reason that the limit on the spin nucleon  cross section extracted from both
targets is much poorer.

The factors $c19= c_{coh}(19,\mu_r(19))$,  $s19= c_{spin}(19,\mu_r(19))$,
$c19= c_{coh}(73,\mu_r(73))$,  $s73= c_{spin}(73,\mu_r(73))$
 and
$c127=c_{coh}(127,\mu_r(127))$,  $s127= c_{spin}(127,\mu_r(127))$
for two values of $Q_{min}$ and $s3= c_{spin}(3,\mu_r(3))$ for  $Q_{min}=0$ can be found elsewhere \cite{JDV06}.
\section{Bounds on the scalar proton cross section}
\begin{figure}
\begin{center}
\rotatebox{90}{\hspace{0.0cm} $\sigma_p\rightarrow 10^{-5}$pb}
\includegraphics[height=.18\textheight]{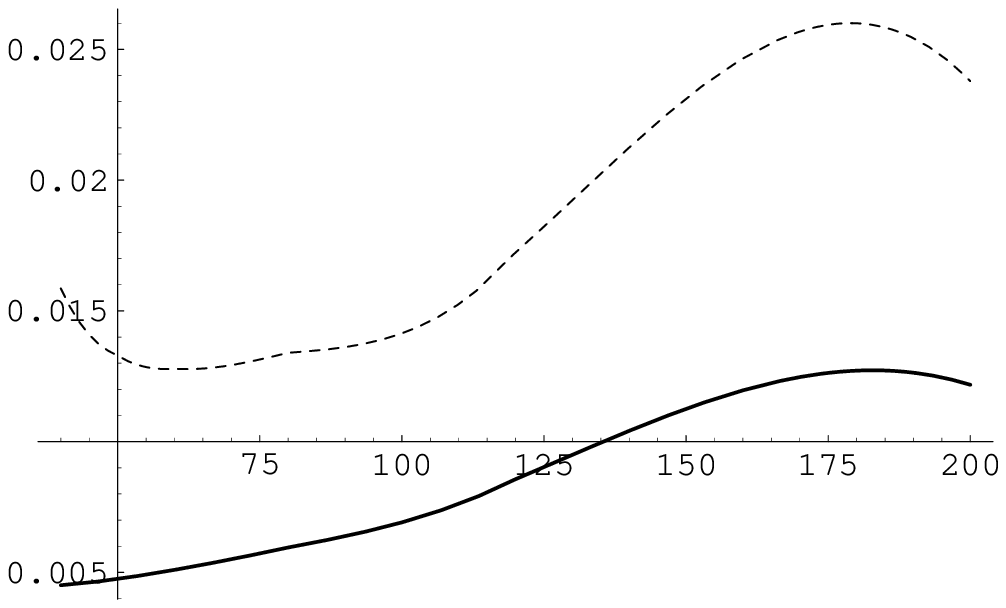}
\hspace{-2.0cm} $m_{\chi}\rightarrow$ GeV
\rotatebox{90}{\hspace{0.0cm} $\sigma_p\rightarrow 10^{-5}$pb}
\includegraphics[height=.18\textheight]{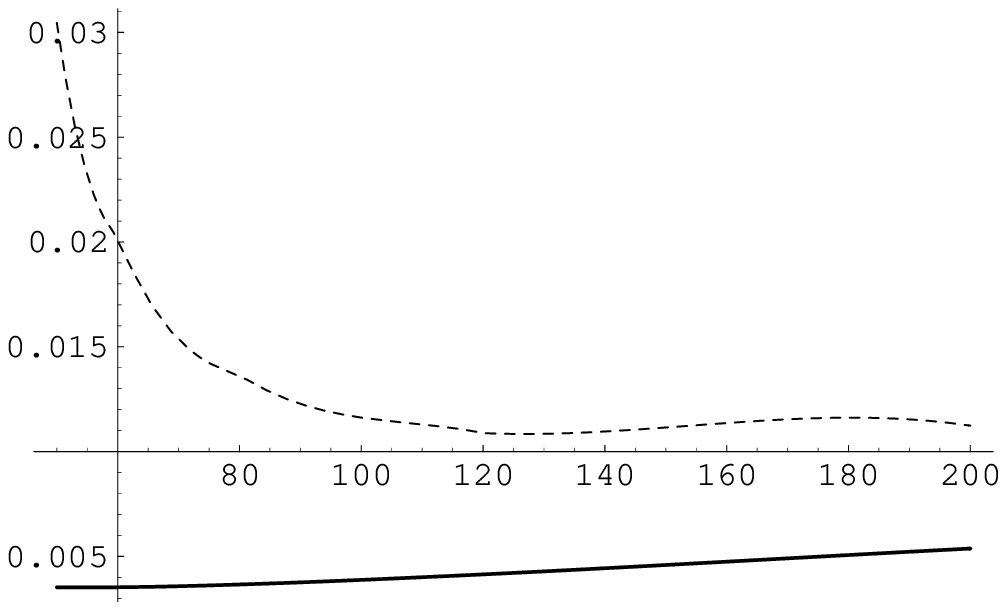}
\hspace{-2.0cm} $m_{\chi}\rightarrow$ GeV
\caption{ The limits on the scalar proton cross section for A$=127$ on the left and A$=73$ on the right as functions of $m_{\chi}$. The continuous (dashed) curves correspond to $Q_{min}=0~(10)$ keV respectively. Note that the advantage of the larger nuclear mass number of the A$=127$ system is counterbalanced by the favorable form factor dependence of the A$=73$ system.
 \label{b127.73}}
\end{center}
\end{figure}

Before proceeding with the analysis of the spin contribution we would like to discuss the limits on the
scalar proton cross section.
In what follows we will employ for all targets \cite{BCFS02}-\cite{PAVAN01} the limit of CDMS II for the Ge target \cite{CDMSII04},
 i.e.  $<2.3$ events for 
an exposure of $52.5$ Kg-d with a threshold of $10$ keV. This event rate is similar to that for other systems \cite{SGF05}. The thus obtained limits are exhibited in Fig. \ref{b127.73}.
 \section{Exclusion Plots in the $a_p,a_n$ and $\sigma_p,\sigma_n$ Planes}
\label{exclplots}
\begin{figure}
\begin{center}
\rotatebox{90}{\hspace{0.0cm} $a_n\rightarrow 0.65$}
\includegraphics[height=.16\textheight]{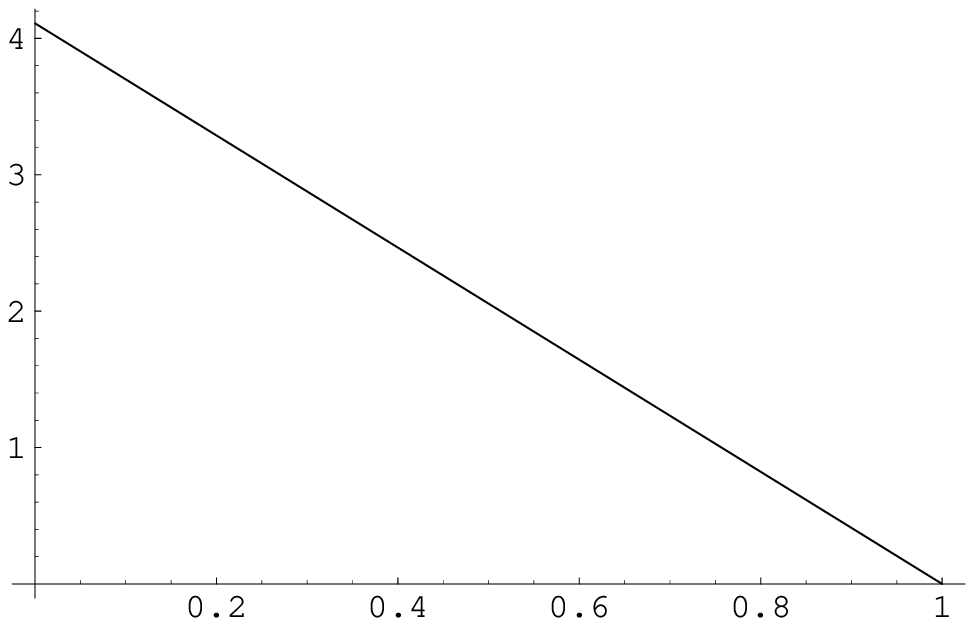}
\hspace{-1.5cm} $a_p\rightarrow 0.65$
\rotatebox{90}{\hspace{0.0cm} $a_n\rightarrow 0.65$}
\includegraphics[height=.16\textheight]{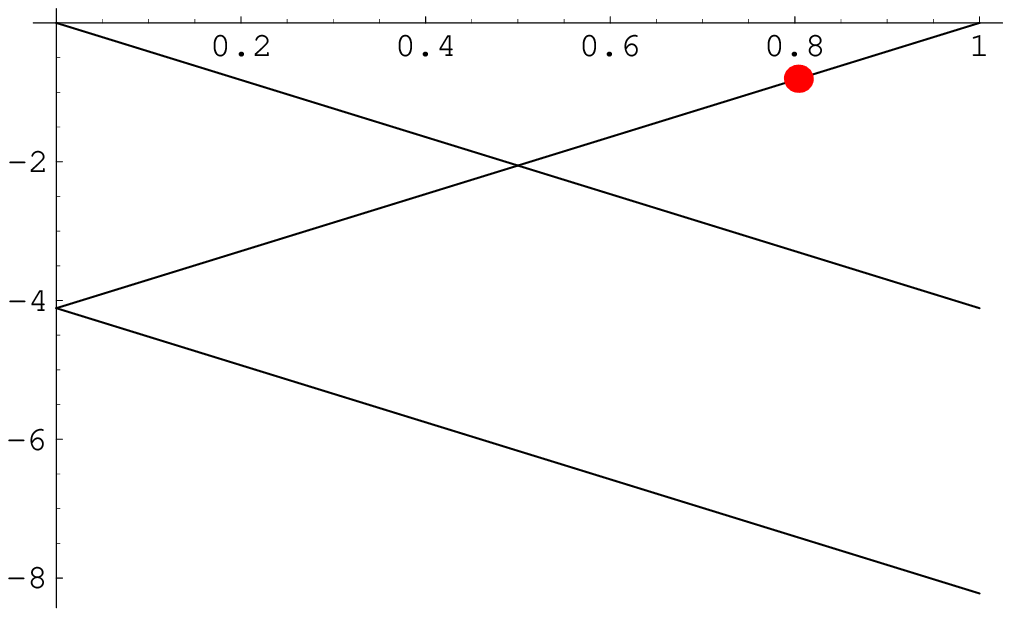}
\hspace{-1.5cm} $a_p\rightarrow 0.65$ \caption{ The boundary in
the $a_p,a_n$ plane extracted from the data for the target
$^{127}I$ is shown assuming  that the amplitudes are relatively
real.  The scale depends on the event rate and the LSP  mass.
Shown here is the scale for $m_{\chi}=100$ GeV.  Note that the
allowed region is confined when the amplitudes are of the same
sign (left plot), but they are not confined when the amplitudes
are of opposite sign. The allowed space now is i) The small
triangle and ii) The space between the two parallel lines and on
the right of the line that intercepts them. We also indicate by a
dot the point $a_p=-a_n$ favored by the spin structure of the
nucleon. The nuclear ME employed were those of Ressel and Dean
(see table \ref{table.spin})
 \label{gpgnamp} }
\end{center}
\end{figure}

\begin{figure}
\begin{center}
\rotatebox{90}{\hspace{0.0cm} $\sigma_n \rightarrow 5.0\times
10^{-3}$ pb}
\includegraphics[height=.16\textheight]{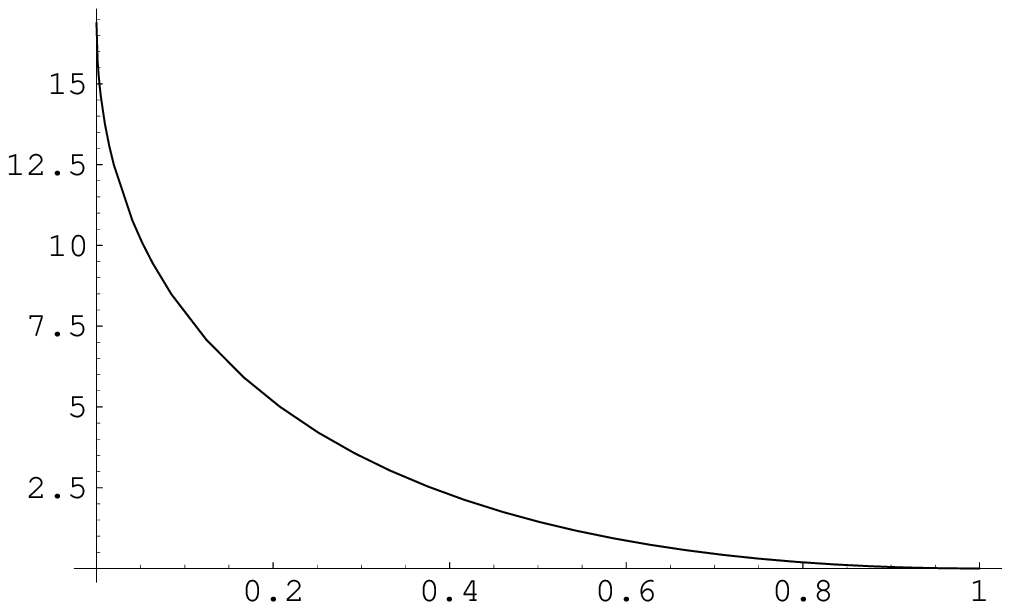}
\rotatebox{90}{\hspace{0.0cm} $\sigma_n\rightarrow 5.0\times
10^{-3}$ pb }
\includegraphics[height=.16\textheight]{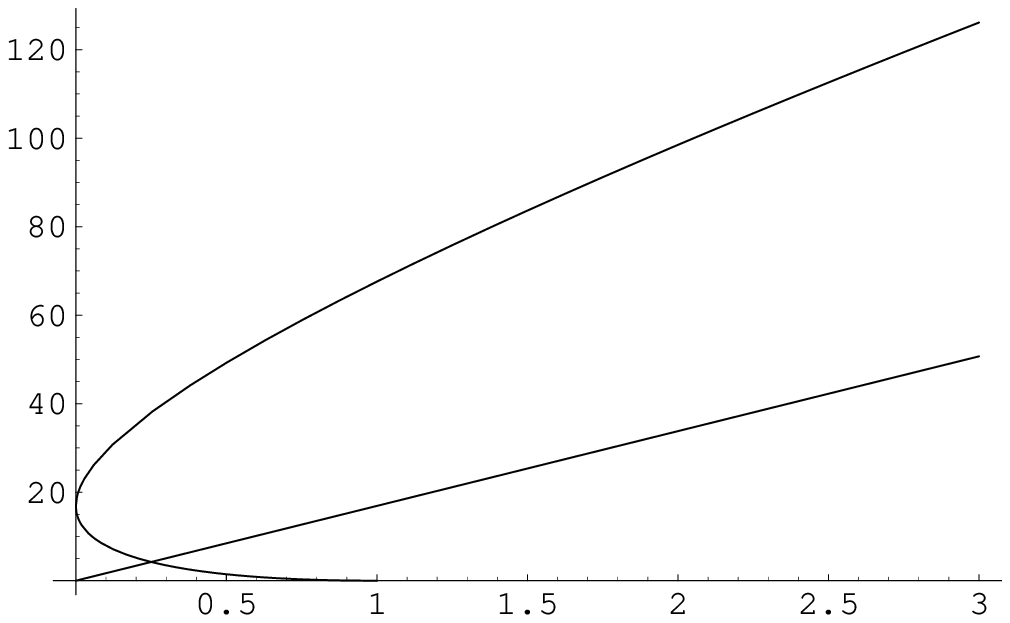}
\hspace{-2.5cm}\\
\hspace{-0.5cm} $\sigma_p\rightarrow 5.0\times 10^{-3}$ pb
\caption{ The same as in Fig. \ref{gpgnamp}  for the
$\sigma_p,\sigma_n$ plane.
 On the left the allowed region  is that below the curve (the amplitudes are relatively real  and
 have the same sign) .  In the plots on the right the amplitudes are relatively real and of opposite sign. The allowed region is i) between the higher segment of the hyperbola and the straight line and ii) Between the straight line and the lower segment of the curve.
 The
nuclear ME employed were those of Ressel and Dean (see table
\ref{table.spin})
 \label{gpgnsigma} }
\end{center}
\end{figure}

\begin{figure}
\begin{center}
\rotatebox{90}{\hspace{0.0cm} $\sigma_n\rightarrow 5.0\times
10^{-3}$ pb}
\includegraphics[height=.2\textheight]{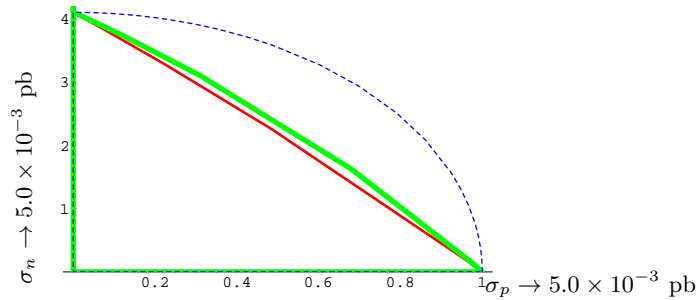}
\hspace{-0.5cm} $\sigma_p\rightarrow 5.0\times 10^{-3}$ pb
\caption{ The same as in Fig. \ref{gpgnsigma}  assuming that the
amplitudes are not relatively real, but are characterized by a
phase difference $\delta$. The allowed space is now confined. The
results shown for the thin solid, thick solid and dashed curves
correspond  to $\delta=\pi /3,~\pi /6$ and $\pi /2$ respectively .
 \label{sigmadelta} }
\end{center}
\end{figure}

 From the data one can extract a
restricted region in the  $\sigma_p,\sigma_n$ plane, which depends
on the event rate and the LSP mass. Some such exclusion plots have already appeared \cite{SGF05}-\cite{GIUGIR05}.
One can plot the constraint imposed on the quantities $|a_p+\frac{\Omega_n}{\Omega_p}a_n|$ and$|\sqrt{\sigma_p}+\frac{\Omega_n}{\Omega_p} \sqrt{\sigma_n}
 e^{i \delta}|^2$ derived from the experimental limits via relations:
 \barr
 &|&\sqrt{\sigma_p}+\frac{\Omega_n}{\Omega_p} \sqrt{\sigma_n}
 )e^{i \delta}|^2\preceq \sigma _{bound}(A)~r(m_{\chi},A),\\
\nonumber
& &\sigma _{bound}(A)=\frac{R}{\bar{K}} \frac{3}{\Omega^2_p}\frac{10^{-5} pb}{c^{100}_{spin}(A,\mu_r(A))}
~,~r(m_{\chi},A)=\frac{c^{100}_{spin}(A,\mu_r(A))}{c_{spin}(A,\mu_r(A))}
 \label{constraintsigma}
 \earr
 where $\delta$ is the phase difference between the two amplitudes and ${c^{100}_{spin}(A,\mu_r(A))}$ is the value of ${c_{spin}(A,\mu_r(A))}$ evaluated for the LSP mass of $100$ GeV. Furthermore
 \beq|a_p+\frac{\Omega_n}{\Omega_p}a_n|\preceq  a _{bound}(A)~\left [r(m_{\chi},A) \right ]^{1/2}~,~a _{bound}(A)=\left[\frac{\sigma _{bound}(A)}{3 \sigma_0} \right]^{1/2}
 \label{constraintampl}
 \eeq
 The constraints will be obtained using the functions ${c^{100}_{spin}(A,\mu_r(A))}$, obtained without
 energy cut off , $Q_{min}=0$, even though the experiments have energy cut offs greater than zero.
Furthermore even though we know of no model such that $e^{i\delta}$ is complex, for completeness we will examine below this case as well. Such plots depend on the relative magnitude of the spin matrix elements. They will be given in units of
the A-dependent quantity $\sigma _{bound}(A)$ for the nucleon cross sections and the dimensionless
quantity $a _{bound}$ for the amplitudes respectively.
 Before we proceed further we should mention  that, if both protons and neutrons contribute, the
standard exclusion plot, must be replaced by a sequence of plots, one for
each LSP mass or via three dimensional plots. We found it is adequate to provide one such plot for a standard LSP mass, e.g. $100$ GeV, and zero energy threshold. The interested reader can find the scale for any other case in work already published \cite{JDV06}.
The situation is exhibited in  Figs \ref{gpgnamp}-\ref{sigmadelta}
in the interesting case of the A=127 system using the nuclear
matrix elements of Ressel {\it et al} given in Table
\ref{table.spin}. For other targets we refer to the literature \cite{JDV06}.
\\One can understand the asymmetry in the plot due
to the fact that $\Omega_p$ is much larger than $\Omega_n$. In
other words if $\sigma_p$ happens to be  very small a large
$\sigma_n$ will be required to accommodate the data.  In the
example considered here, however, the extreme values differ only
by $20\%$
 from the values on the axes,
which arise, if one assumes that one mechanism at a time (proton or neutron) dominates.
\section{The modulation effect.}
\begin{figure}
\begin{center}
\rotatebox{90}{\hspace{1.0cm} $h\rightarrow$}
\includegraphics[height=0.15\textheight]{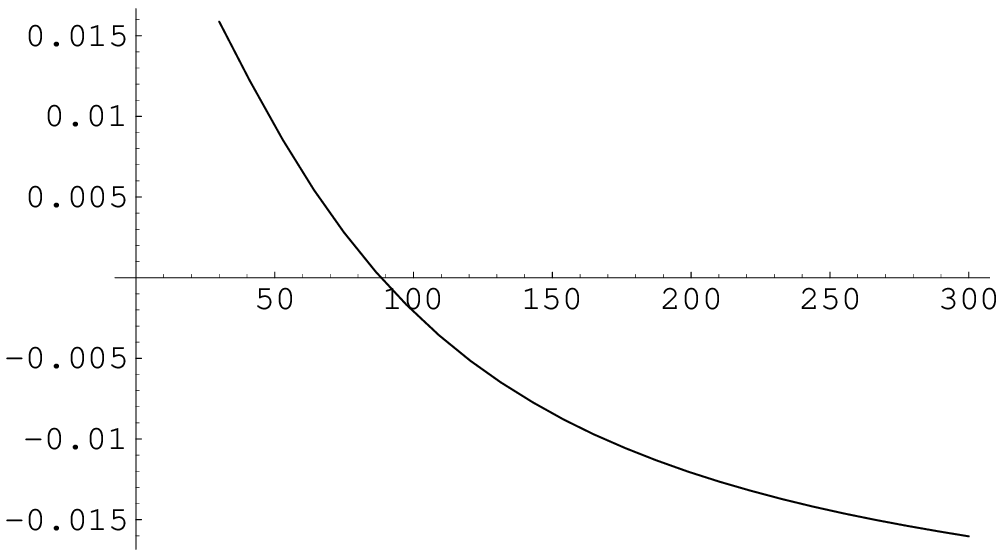}
\rotatebox{90}{\hspace{1.0cm} $h\rightarrow$}
\includegraphics[height=0.15\textheight]{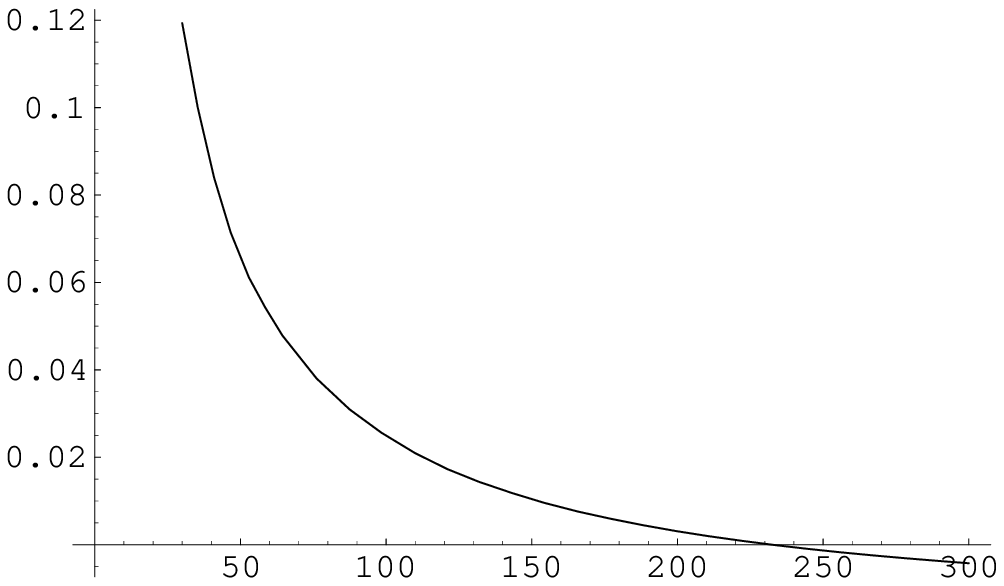}
 \hspace{0.cm}\\
 \hspace{0.5cm} $m_{\chi}\rightarrow$ ($GeV$)
\caption{ The modulation
amplitude $h$ as a function of the LSP mass in the case of ${127}$I for $Q_{min}=0$ on the left
and $Q_{min}=10$ keV on the right. We should mention that the average LSP energy for an LSP mass $m_{\chi}=100$ 
GeV is $\simeq 40$ keV.
  For the definitions see text.
\label{hgs} }
\end{center}
\end{figure}

As we have mentioned the expected event rate is so low that, even if one goes underground, the background is formidable. Especially since the signal coming from  the detection of the energy energy of the recoiling nucleus has the same shape as that of the background. One, therefore, looks for specific signatures associated with the reaction. Since the event rate depends on the relative velocity between the LSP and the target, a periodic seasonal dependence is expected due to the motion of the Earth around the sun. What counts is the the is the projection of the velocity of the earth on the sun's velocity (see Fig. \ref{fig:velocity}).
\\If the effects of the motion of the Earth around the sun are included, the total
 non directional rate is given by
\begin{equation}
R= \bar{K} \left[c_{coh}(A,\mu_r(A)) \sigma_{p,\chi^0}^{S}(1 +  h(a,Q_{min})cos{\alpha})\right]
\label{3.55j}
\end{equation}
and an analogous one for the spin contribution.  $h$  is the modulation amplitude, which is quite small, less than
$2\%$ and it depends on the velocity distribution, the nuclear form factor and, for a given target, on the LSP mass.
 $\alpha$ is the phase of the Earth, which is
zero around June 2nd. In the case of the target $^{127}$I the modulation amplitude is shown in Fig. \ref{hgs}.
We see that the modulation amplitude is small, especially for $Q_{min}=0$. Furthermore its sign is uncertain, since it depends on the LSP mass. The modulation amplitude increases as the threshold cut off energy increases, but, unfortunately, this occurs at the expense of the total number of counts. Furthermore many experimentalists worry that there are may be seasonal variations in the relevant backgrounds as well.
\section{Transitions to excited states}
 As we have mentioned the average neutralino energy scales with its mass. It is $\simeq 40$ keV for $m_{\chi}=100$ GeV. Thus the neutralino energy is not high enough to excite the nucleus. In some rare cases involving odd mass nuclei there exist excited states at low energies, which can be populated in the LSP-nucleus collision due to the high velocity tail of the neutralino velocity distribution. From an experimental point of view this is very interesting \cite{eji93}, since the signature of the $\gamma-$ray emission following the nuclear de-excitation is much easier than nuclear recoils. An interesting target is $^{127}$I, which has an excited state at $\simeq 50$ keV. It has recently been found \cite{VQS04} that the branching ratio to this excited state is appreciable from an experimental point of view.
 
\begin{figure}
\begin{center}
\rotatebox{90}{\hspace{1.0cm} {\tiny BRR}$\rightarrow$}
\includegraphics[height=.18\textheight]{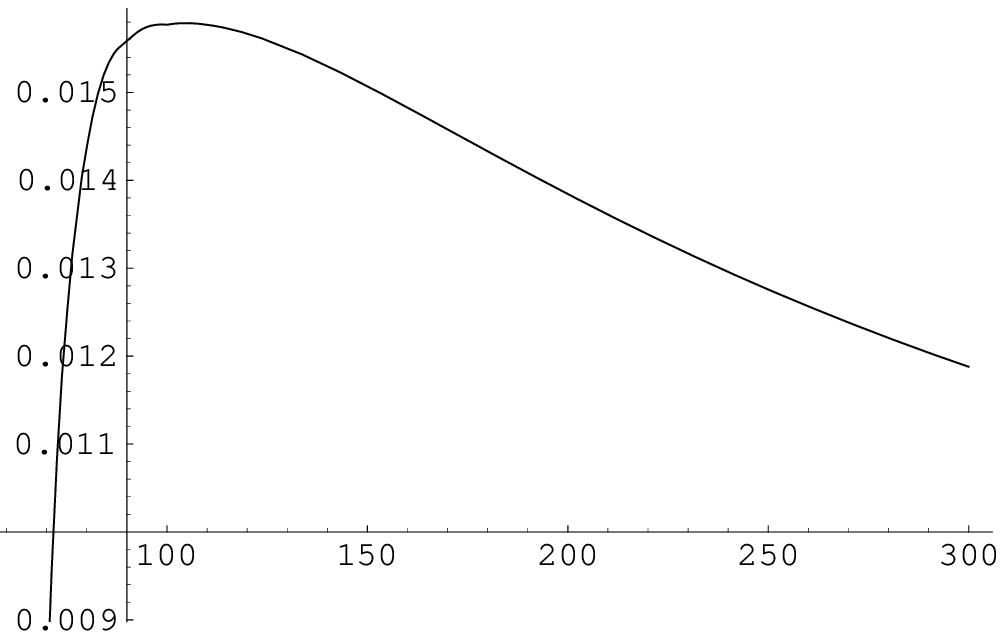}
 \rotatebox{90}{\hspace{1.0cm} {\tiny BRR}$\rightarrow$}
\includegraphics[height=.18\textheight]{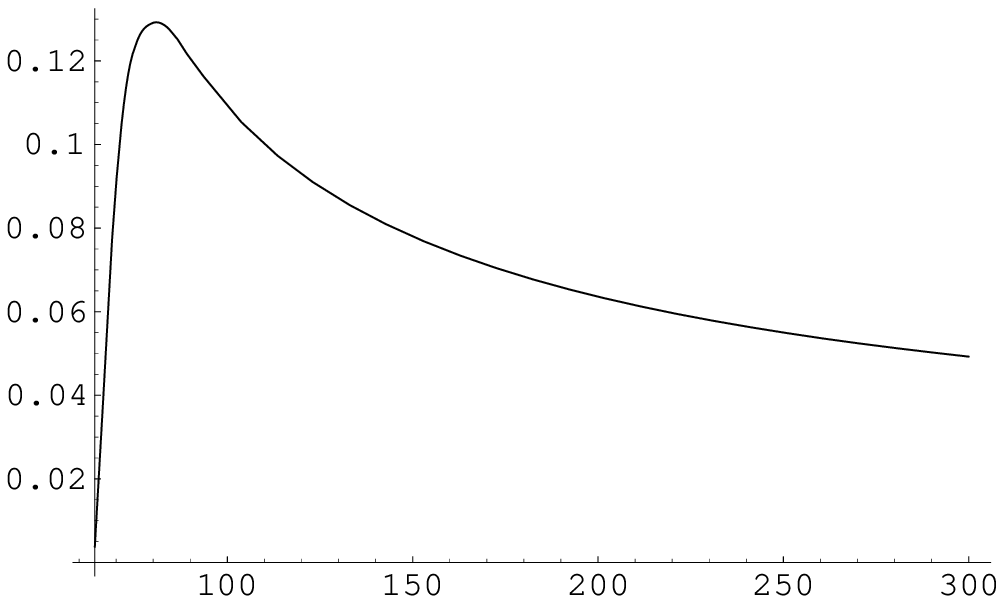}
\hspace{0.0cm}\\
 \hspace{0.0cm}$m_{\chi}\rightarrow$ ($GeV$)
 \caption{ The ratio of the rate to
the excited state divided by that of the ground state as a
function of the LSP mass (in GeV) for $^{127}I$. It was found that
the static spin matrix element of the transition from the ground
to the excited state is a factor of 1.9 larger than that
involving the ground state. The spin response functions
$F_{11}(u)$ were assumed to be
the same. On the left we show the results for $Q_{min}=0$ and on
the right for $Q_{min}=10~KeV$. In the last case, due to the detector energy cut, off the denominator (recoil rate) is reduced, while the numerator (the rate  to the excited state) is not affected. 
 \label{ratio} }.
\end{center}
\end{figure}
\section{The directional rates}
\label{directional}
\begin{figure}
\begin{center}
\rotatebox{90}{\hspace{1.0cm} $t_{dir}\rightarrow$}
\includegraphics[height=.15\textheight]{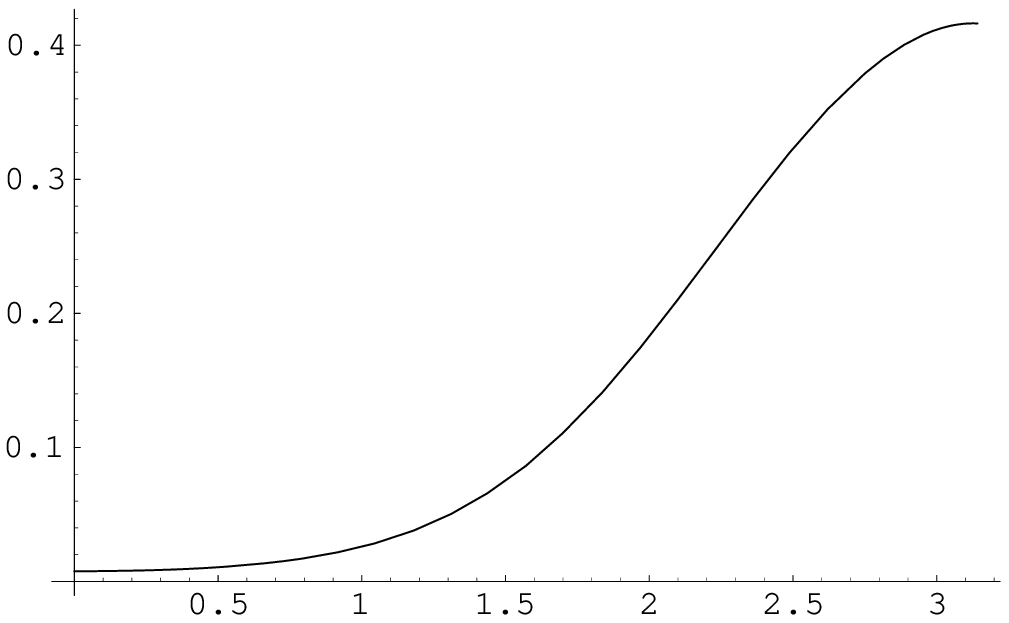}
\hspace{0.0cm} $\Theta \rightarrow$
\rotatebox{90}{\hspace{1.0cm} $t_{dir}\rightarrow$}
\includegraphics[height=.15\textheight]{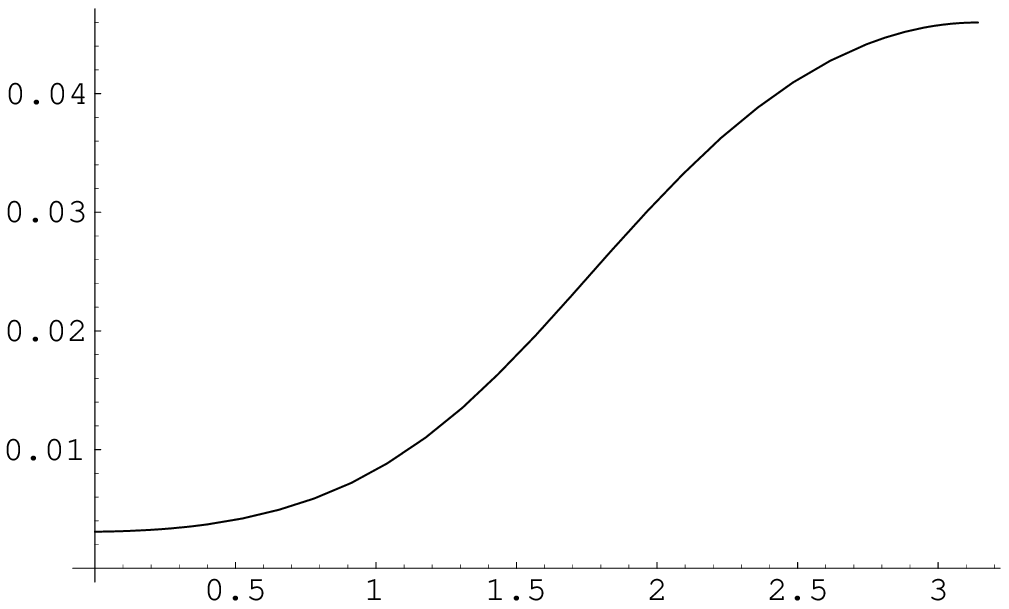}
\hspace{0.0cm} $\Theta \rightarrow$ 
 \caption{ The quantity $\kappa t$
as a function of the angle $\Theta$, the polar angle from the sun's direction of motion,
for $A=19$ on the left and $A=127$ on the right. 
The results presented correspond to an LSP mass of $100~GeV$.
 \label{tdir}
 }
\end{center}
\end{figure}

 As we have already mentioned one may attempt to measure not only the energy of the recoiling nucleus, but observe its direction of recoil. Admittedly such experiments are quite hard \cite{DRIFT}, but they are expected to provide unambiguous signature against  background rejection.
 Since the sun is moving around the galaxy in a directional experiment, i.e. one in which the
direction of the recoiling nucleus is observed, one expects a strong correlation of the
event rate with the motion of the sun \cite{JDVSPIN04}. In fact
the directional rate can be written as:
\begin{equation}
R_{dir} = \frac{\kappa} {2 \pi} \bar{R}~t  \,
            [1 + h_m  cos {(\alpha-\alpha_m~\pi)}]
\label{4.56b}
\end{equation}
where  $h_m$ is the modulation and
 $\alpha_m $ is the "shift" in the phase of the Earth $\alpha$,
 since now  the maximum
 occurs at $\alpha=\alpha_m \pi$. $\kappa/(2 \pi)$ is the reduction
 factor of the unmodulated
directional rate relative to the non-directional one. The
parameters  $\kappa~,~h_m~,~\alpha_m$ depend on the direction of
 observation:
$$\hat{e}=(\sin{\Theta} \cos{\Phi} ~,~\sin{\Theta} \sin{\Phi}~,~\cos{\Theta})$$
 The parameter $\kappa t$ for a typical LSP mass $100~GeV$ is shown in
 Fig. \ref{tdir} as a function of the angle $\Theta$ for the targets $A=19$
 and $A=127$.
 We see that the change of the rate as a function of
 the angle $\Theta$ for the Maxwellian LSP velocity distribution is quite
 dramatic. This figure is  important in the analysis of the angular
 correlations, since, among other things, there is always un uncertainty in
 the determination of the angle in a directional experiment.
We  prefer to use the parameters
$\kappa$ and $h_m$, since, being ratios, are
expected to be
 less dependent on the parameters of the theory. We
 exhibit the dependence of the
parameters $t$, $h$, $\kappa,h_m$, and $\alpha_m$,
which are essentially independent of
the LSP mass  for target $A=19$, in
 Table \ref{table1.gaus} (for the other light systems the results are
almost identical).

The asymmetry is quite large. For a Gaussian velocity distribution we find:
$$As=\frac{R(-z)-R(+z)}{R(-z)+R(+z)}\approx 0.97$$ In the other directions
it depends on the phase of the Earth and is equal to almost twice
 the modulation.
For a heavier nucleus the situation is a bit complicated. Now the
parameters $\kappa$ and $h_m$ depend on the LSP mass \cite{JDVSPIN04}. It is clear that, if
such experiments will ever be performed, such signatures  cannot be mimicked by background events.
\begin{table}[t]
\caption{ The parameters $t$, $h$, $\kappa,h_m$ and $\alpha_m$ for the
 isotropic Gaussian
 velocity distribution and $Q_{min}=0$. The results presented are  associated
 with the spin contribution, but those
 for the coherent mode are similar. The results shown are for the light
systems. For intermediate and heavy nuclei there is a dependence on the LSP mass. $+x$ is
 radially  out of the galaxy ($\Theta=\pi/2,\Phi=0$), $+z$ is in the sun's
 direction of motion ($\Theta=0$) and
$+y$ is vertical to the plane of the galaxy ($\Theta=\pi/2,\Phi=\pi/2$) so that
 $(x,y,z)$ is right-handed. $\alpha_m=0,1/2,1,3/2$ means
that the maximum occurs on the 2nd of June, September, December and March
 respectively.
\label{table1.gaus}}
\begin{center}
\begin{tabular}{lrrrrrr}
& & & & & &      \\
type&t&h&dir &$\kappa$ &$h_m$ &$\alpha_m$ \\
\hline
& & & & & &      \\
& &&+z        &0.0068& 0.227& 1\\
dir& & &+(-)x      &0.080& 0.272& 3/2(1/2)\\
& & &+(-)y        &0.080& 0.210& 0 (1)\\
& & &-z         &0.395& 0.060& 0\\
\hline
all&1.00& & && & \\
all& & 0.02& & & & \\
\hline
\end{tabular}
\end{center}
\end{table}
\section{Observation of electrons produced during the LSP-nucleus collisions}
\begin{figure}
\includegraphics[height=6.0cm,width=6.0cm]{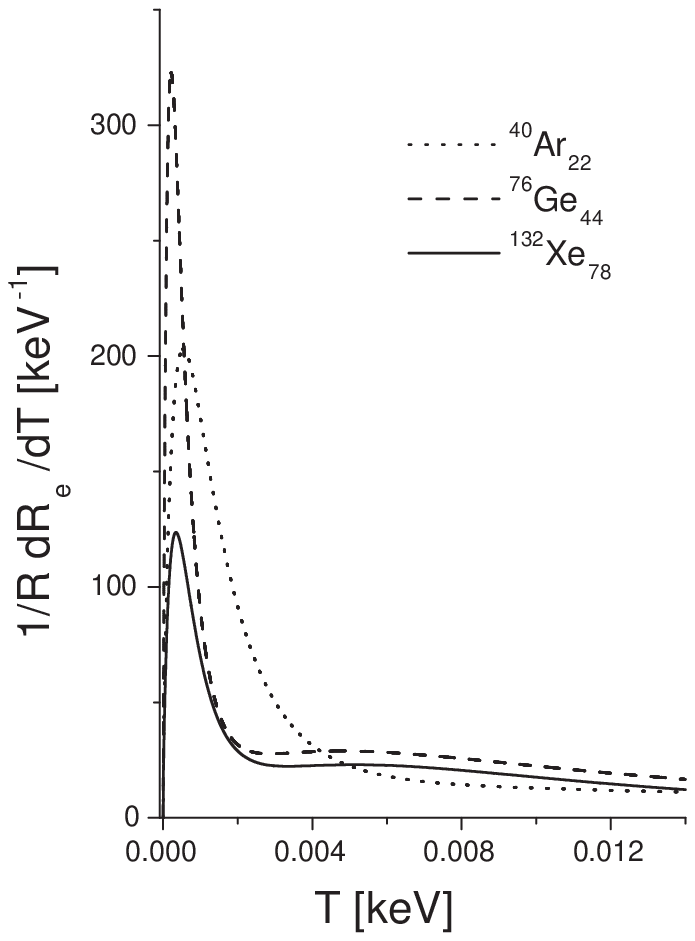}
\
 \includegraphics[height=6.0cm,width=6.0cm]{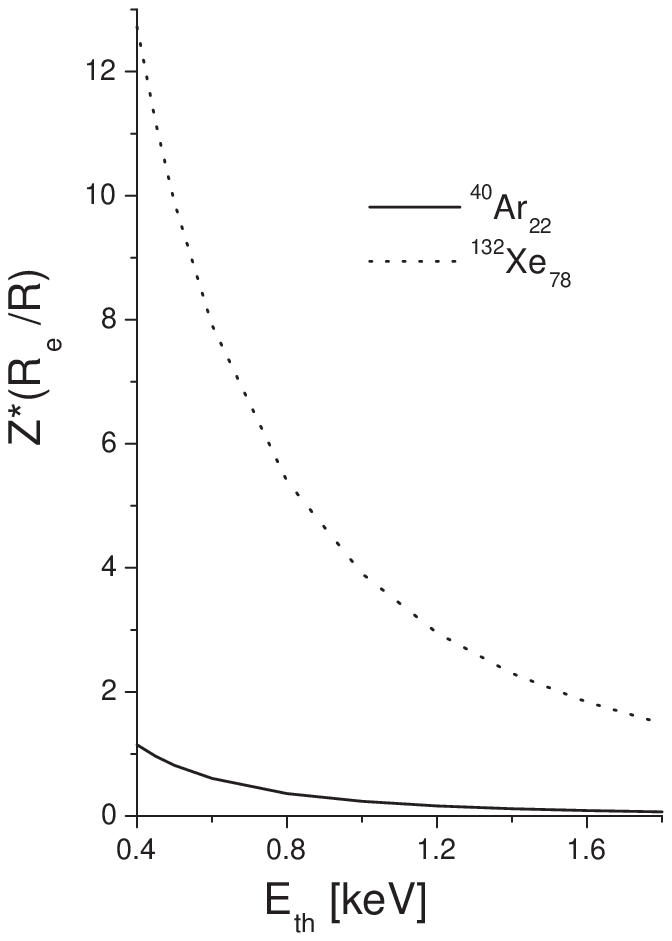}
 \caption{ On the left we show the differential rate for ionization electrons, divided by the total rate associated
  with the nuclear recoils, as a function of the electron energy $T$ (in keV) for
  various atoms. On the right we show  the total rate for producing electrons divided by the corresponding rate for nuclear recoil as a function of the threshold energy. The event rate is per atom, i.e. all electrons in the atom have been considered.  The results exhibited were obtained for
  a typical LSP mass $m_{\chi}=100$ GeV. }
\label{zfig}.
\end{figure}

 Since the detection of recoiling nuclei is quite hard one may look for other events. One such possibility is the observation of ionization electrons produced directly during the LSP nuclear collisions \cite{VE05}, \cite{MVE05}. Due to the properties of the bound electron wf, the event rate peaks at very low electron energies. One therefore must be able to achieve very low energy thresholds. In order to avoid uncertainties arising from the constraint SUSY parameter space we have opted to present the ratio of the event rate for producing electrons divided by the standard coherent recoil rate. This ratio is exhibited as a function of the electron threshold energy in Fig. \ref{zfig}. We see that for large atomic number Z and sufficiently low threshold energy this ratio may exceed unity.
 
 It has also been found that inner 1s electrons can be ejected with a non negligible probability \cite{EMV05}. The produced electron holes can be filled via the Auger process or a sizable fraction can proceed via very hard (32 keV) X-ray emission. The detection of such X-rays, in or without coincidence with nuclear recoils, will provide a signature very hard to miss, if SUSY allows for detectable recoil rates.
\section{Conclusions}
 In this review we have dealt with various issues involving the direct detection of supersymmetric dark matter. The standard experiments employ various techniques of measuring the energy of the recoiling nuclei after their elastic scattering with the dark matter candidates. We have seen that the evaluation of the event rates  involves a number of issues: 1) A supersymmetric model with a number of parameters, which at present can only be constrained from laboratory data at low energies as well as cosmological observations. 2) The dependence of the nucleon cross section on quarks other than u and d. 3) A proper nuclear model, which involves the nuclear form factor in the case of the the scalar interaction and the spin response function for the axial current. 4)  Information about the density and the velocity distribution of the neutralino (halo model).\\
 Using the present experimental limits on the event rate and suitable inputs in 3)-4) we have derived constraints in the nucleon cross sections. Since the obtained event rates are extremely low, we have examined some additional signatures inherent in the neutralino nucleus interaction, such as the periodic behavior of the rates due to the motion of Earth (modulation effect). Since, unfortunately, this is characterized by a small amplitude, we were lead to examine the possibility of directional experiments. Tese, in addition to the recoil energy, will also attempt to measure the direction of the recoiling nuclei. The event rate in a given direction is $\sim 6\pi$ smaller than that of the standard experiments, but one maybe able to  exploit two novel characteristic signatures: a) large asymmetries and b) interesting modulation patterns.\\
 Proceeding further we extended our study to include evaluation of the rates for other than recoil searches such as: i) Transitions to excited states and the observation of de-excitation $\gamma$ rays, ii) detection of the recoiling electrons produced during the neutralino-nucleus collision and iii) observation of hard X-rays, following the de-excitation of the ionized atom.\\
With all the above signatures one hopes that, if the supersymmetric models do not conspire to lead  to large suppression of the amplitudes, the direct direction of dark matter may soon follow. 

\section*{Acknowledgements}
This work was supported by European Union under
the contract MRTN-CT-2004-503369 as well as the program PYTHAGORAS-1. The latter is part of the
Operational Program for Education and Initial Vocational Training of the
Hellenic Ministry of Education under the 3rd Community Support Framework
and the European Social Fund. The author is indebted to Professor Lefteris Papantonopoulos for support and hospitality during  the Aegean Summer School.
 
\end{document}